\documentclass[12pt,peerreview]{IEEEtran0}
\usepackage{cite}
\usepackage{graphicx}
\usepackage{subfigure}
\usepackage{fancyhdr}
\usepackage{url}
\usepackage[centertags]{amsmath}
\usepackage{amsmath,amssymb,amstext}
\usepackage{amsfonts}
\usepackage{stfloats}
\usepackage{algpseudocode}
\usepackage{algorithm}
\newcounter{saveeqn}%

\begin{document}

\title{Joint Downlink Cell Association and Bandwidth Allocation
for Wireless Backhauling  in Two-Tier HetNets with Large-Scale Antenna Arrays}
%\author{Ning Wang,~{\it Member,~IEEE,}
%        Ekram Hossain,~{\it Fellow,~IEEE,}\\
%        and~Vijay~K.~Bhargava,~{\it Life~Fellow,~IEEE}
\author{Ning Wang, Ekram Hossain, and Vijay K. Bhargava
\thanks{N. Wang and V. K. Bhargava are with the Department of Electrical and Computer Engineering,
The University of British Columbia, Vancouver, BC, Canada V6T 1Z4 (email: \{ningw, vijayb\}@ece.ubc.ca).}
\thanks{E. Hossain is with the Department of Electrical and Computer Engineering,
University of Manitoba, Winnipeg, MB, Canada R3T 5V6
(e-mail: ekram.hossain@umanitoba.ca).}
\thanks{Preliminary results of this work have been submitted to the {\em 28th Annual
IEEE Canadian Conference on Electrical and Computer Engineering (CCECE'2015)}.}}

%\IEEEoverridecommandlockouts

\setcounter{page}{1}
\maketitle
\thispagestyle{empty}\pagestyle{empty}
%\IEEEpeerreviewmaketitle

\begin{abstract}
The problem of joint downlink cell association (CA) and wireless backhaul bandwidth allocation (WBBA)
in two-tier cellular heterogeneous networks (HetNets) is considered.
Large-scale antenna array is implemented at the macro base station (BS),
while the small cells within the macro cell range are single-antenna BSs
and they rely on over-the-air links to the macro BS for backhauling.
%The wireless backhaul constraint becomes another key criterion for cell association
%and radio resource allocation.
A sum logarithmic user rate maximization problem is investigated considering wireless backhauling constraints.
A duplex and spectrum sharing scheme based on co-channel reverse time-division duplex (TDD)
and dynamic soft frequency reuse (SFR) is proposed for interference management in two-tier HetNets
with large-scale antenna arrays at the macro BS and wireless backhauling for small cells.
%Because no additional radio frequency hardware is required by the proposed scheme,
%it allows low-cost and fast deployment of wireless backhaul enabled 5G cellular HetNets.
Two in-band WBBA scenarios, namely, unified bandwidth allocation and per-small-cell bandwidth allocation scenarios,
are investigated for joint CA-WBBA in the HetNet.
A two-level hierarchical decomposition method for relaxed optimization is employed
to solve the mixed-integer nonlinear program (MINLP).
Solutions based on the General Algorithm Modeling System (GAMS) optimization solver and fast heuristics
are also proposed for cell association in the per-small-cell WBBA scenario.
It is shown that when all small cells have to use in-band wireless backhaul,
the system load has more impact on both the sum log-rate and per-user rate performance
than the number of small cells deployed within the macro cell range.
The proposed joint CA-WBBA algorithms have an optimal load approximately equal to the size of
the large-scale antenna array at the macro BS.
The cell range expansion (CRE) strategy, which is an efficient cell association scheme for HetNets
with perfect backhauling, is shown to be inefficient when in-band wireless backhauling for small cells comes into  play.
\end{abstract}

%\vspace{-15mm}


%%%%%%%%%%%%%%%%%%%%%%%%%%%%%%%%%%%%%%%%%%%%%%%%%%%%%%%%%%%%%%%%%%%%%%%%%%%%%%%%%%%%%%%%%%%%%%%
\section{Introduction}\label{one}

%The exponentially increasing mobile IP traffic due to the rapid growth of
%smart portable devices and the associated multimedia and cloud-based services
%has become a major driving force for the worldwide deployment of the 3GPP Long Term Evolution (LTE)
%and a demand for the next generation cellular technology, i.e., 5G.
%Because of the scarce of the radio spectra suitable for mobile communications,
%base station densification is an inevitable technological choice for future cellular systems, which will improve
%area spectral efficiency such that the traffic growth can be accommodated with the limited bandwidth.

In the upcoming 5G cellular standards, the radio access network (RAN) is going to experience a paradigm shift
rather than an incremental upgrade of the 3G and LTE RAN \cite{hetnetshift}.
Conventional cell splitting for single-tier cellular architecture will not be suitable
because of the extraordinarily complex radio network planning for interference management,
and a huge demand for additional backhaul resources.
The multi-tier heterogeneous network (HetNet) architecture, incorporated with flexible backhaul connections
and advanced physical layer technologies such as large-scale multi-user multiple-input-multiple-output (MIMO)
antenna systems, is a promising solution \cite{5gimp}.
With a large number of small cells (e.g., pico and femto cells) deployed
 inside the macro cell range to meet the capacity and coverage demands, the network topology will see a substantial change.
The newly incorporated technological and architectural enhancements
will impose additional constraints to the system design.
In particular, cell association (CA) and radio resource allocation mechanisms will be fundamentally different,
and they will have great impact on the performance of 5G cellular networks.

Limitations in the backhauling systems for small cells has yet been adequately considered
in previous studies of HetNets \cite{bennis}.
With a very dense deployment of small cells, it is impractical to have wired connections at every small cell site. Instead, in-band wireless backhauling, which allows low-cost and fast deployment of a new cellular network, is a desirable solution.
By implementing plug-and-play small cell equipments which can use in-band over-the-air links
to macro base stations for backhauling,
the operators may scale up their multi-tier 5G cellular networks in a short time.
However, the use of in-band wireless backhauling gives rise to additional source of interference,
which subsequently impacts the cell association and resource allocation strategies.
It is therefore important to consider various backhauling scenarios in the context of designing cell association and resource allocation (or interference management) methods.
For improved interference control under wireless backhauling, sophisticated technologies,
e.g., large-scale MIMO, should be considered.

%Particularly, here we study a system employing the large-scale MIMO technology \cite{scaleupmimo},
%which can be used to achieve high-order multiplexing and multiple access by exploiting spatial degrees of freedom.

This work studies the cell association problem in a two-tier HetNet where the small cell tier uses in-band wireless backhauling.
Large-scale MIMO is considered at the macro base station (BS) to mitigate intra- and inter-cell interferences
by exploiting the spatial degrees of freedom \cite{scaleupmimo}.
Specifically, the macro base stations are equipped with large-scale antenna arrays and wired high-capacity backhaul connections.
The single-antenna small cell tier, on the other hand,
relies on in-band over-the-air large-scale MIMO links to the macro cell tier for backhauling.
A macro BS therefore acts as a hub of the in-band wireless backhaul connections for small cells
within its cell range.
In such a system, we can either: i)  dedicate certain amount of bandwidth for in-band wireless backhauling for small cells
throughout the network, or ii) dynamically adjust the bandwidth allocation for backhauling locally at each small cell.
Based on these two in-band small cell wireless backhaul bandwidth allocation (WBBA) strategies,
the corresponding joint CA-WBBA scenarios for the two-tier HetNet are considered and analyzed.
The main contributions of the paper can be summarized as follows.

\begin{itemize}
\item For in-band wireless backhauling for small cells, two bandwidth allocation scenarios,
namely, unified WBBA (u-WBBA) and per-small-cell WBBA (p-WBBA), are investigated for joint CA-WBBA in a two-tier HetNet with large-scale antenna arrays at the macro BS.

\item A duplex and spectrum sharing framework based on co-channel reverse time-division duplex (TDD)
and dynamic soft frequency reuse (SFR) is proposed for interference management in the HetNet
to facilitate the implementation of large-scale MIMO at the macro BS and in-band wireless backhauling for small cells.

\item For the u-WBBA scenario, a relaxed convex optimization approach is employed to
solve the mixed-integer nonlinear program (MINLP) for the joint CA-WBBA problem for maximization of sum of logarithmic user rates. A distributed algorithm based on a two-level hierarchical decomposition framework is proposed.

\item The joint CA-WBBA under the p-WBBA scenario is shown to be a non-linear and non-convex MINLP problem.
The distributed algorithm proposed for the u-WBBA scenarios is extended to the p-WBBA scenario by
considering separate and iterative cell association and WBBA design.
A fast heuristic algorithm of low-complexity is proposed to provide sub-optimal solutions for joint CA-WBBA
in the p-WBBA scenario for online implementations. %with stringent delay requirements.
The General Algorithm Modeling System (GAMS) optimization tool is also employed to solve the sum log-rate maximization problem.
\end{itemize}

The remainder of this paper is organized as follows.
In Section \ref{two}, some closely related works are reviewed.
The system model under investigation is then described in Section \ref{three}.
In Section \ref{four} we study joint CA-WBBA for a two-tier HetNet under the u-WBBA scenario,
and a distributed algorithm is proposed based on relaxation of the cell association variables
and hierarchical decomposition.
Section \ref{five} first extends the distributed algorithm to the p-WBBA case and
proposes a fast heuristic algorithm for macro base station offloading.
Numerical results are presented in Section \ref{six} before the paper is concluded in Section \ref{seven}.

%%%%%%%%%%%%%%%%%%%%%%%%%%%%%%%%%%%%%%%%%%%%%%%%%%%%%%%%%%%%%%%%%%%%%%%%%%%%%%%%%%%%%%%%%%%%%%%%%%
\section{Related Work}\label{two}

\subsection{Cell Association}\label{2a}

The received signal-to-interference-plus-noise ratio (SINR) is an indication
of radio signal quality and is directly related to %symbol error rate (SER) and outage probability
error and outage performance.
It is therefore a natural criterion for cell association and has been employed in most concurrent wireless systems.
However, in multi-tier HetNets,
when the BS load (i.e., number of associated mobile terminals) is taken into consideration,
the SINR-based cell association is in general not optimal.
The macro base stations in a multi-tier HetNet typically have much higher transmit power than the small cells.
Load imbalance will then occur with SINR-based cell association \cite{myths}. That is,
the macro cell is likely to be over-loaded, while the small cells will have much lower load.

As pointed out in \cite{myths}, in a multi-tier HetNet,
optimal cell association is coupled with scheduling and radio resource allocation,
which gives rise to a combinatorial optimization problem.
Large power variations between different tiers make the problem even more involved.
An effective solution is cell range expansion (CRE) of small cells through a biasing mechanism \cite{CRE}.
Even though determination of optimal bias is a challenging problem,
the potential of the CRE for load balancing and system throughput improvement has been recognized.
The almost blank subframe (ABS) for adaptive resource partitioning is a means to eliminate inter-cell
interference when CRE is used \cite{blank}.
The unbiased (macro) cells intentionally leave specific sub-frames blank so that the offloaded users
associated with small cells are scheduled in these subframes to avoid inter-tier interference.
A relaxed optimization approach was employed in \cite{ye} to convert the combinatorial optimization
for cell association into a convex problem, which is mathematically more tractable.
A distributed algorithm, which avoids heavy overhead, was proposed through dual decomposition.
In \cite{hamid}, the quality-of-service (QoS) provisioning of multi-tier HetNets was taken into consideration
for cell association design.
A unified framework for outage minimization and rate maximization  was investigated,
and distributed cell association algorithms were proposed.
Stochastic geometry is another tool, which has recently been used for
modeling and analysis of cell association in multi-tier HetNets \cite{geometry} \cite{geometryBL}.
It models the locations of both the BSs and users as random point processes
and analyzes the system performance statistically.

\subsection{Wireless Backhauling}\label{2b}

In order to satisfy the stringent delay requirements in the future 5G systems \cite{what5g},
it is important to jointly consider the designs of the access network and the backhaul network.
A few recent works in the literature have investigated how limitations of the backhauling system for small cells
affect interference management in multi-tier HetNets.
Inter-tier interference in uplink of femtocell-macrocell overlay HetNets was studied in \cite{shitz},
where unreliable backhaul with varying capacity at femtocells was taken into consideration.
Macro BS offloading problem for the uplink of a HetNet, where small cells rely on capacity-limited
heterogeneous backhaul systems, was considered in \cite{bennis}.
Based on a noncooperative game formulation of the problem,
a distributed offloading algorithm employing a reinforcement learning approach was proposed.
Macro cell users can therefore optimize their performance by being offloaded to
neighbouring small cells.
However, as pointed out in \cite{bennis}, in spite of a rich body of literature on multi-tier HetNets,
there have not been sufficient studies on the impact of small cell backhauling solutions on
radio resource management of HetNets.
Few studies can be found which consider small cell backhauling in the context of designing
cell association schemes for multi-tier HetNets.

In-band wireless backhauling, which uses macro base stations as hubs of the backhaul system
and connects the small cells to the hubs via in-band over-the-air links, would  allow fast and cost-efficient implementations of large-scale 5G networks.
To the best of our knowledge, the problem of cell association in HetNets considering in-band wireless backhauling constraints and large-scale antenna arrays at the macro BS,
 has not been investigated in the literature.

\subsection{Large-scale MIMO}\label{2c}

Large-scale MIMO, also known as massive MIMO, is an emerging multi-user BS technology
that significantly improves radio spectral efficiency
in a system with only single-antenna user devices \cite{massive}.
Large-scale MIMO is considered as a disruptive technology for the future 5G cellular standards \cite{disruptive}.
In 5G multi-tier HetNets, large-scale MIMO can be used to effectively eliminate inter-tier interference.
For example, a hotspot HetNet communication scenario was considered in \cite{adhikary},
and a {\it spatial blanking} strategy of the macro cells equipped with large-scale MIMO was studied
to mitigate inter-tier interference.
By employing a large-scale BS antenna array,
the RF beam can be focused into ever-smaller mobile terminal (MT) spots and thus making
the interference more ``localized'' \cite{scrole}. The
co-channel interference can be effectively eliminated,
and remarkable improvement in spectral efficiency can be achieved.
Consequently, we are able to realize more efficient joint transmission coordinated multipoint (JT CoMP)
and more robust interference suppression.

When small cells in the network have to rely on wireless links to the macro BS
for backhauling, large-scale MIMO is a promising solution which not only improves service to macro cell users,
but also provides good (in-band) wireless backhaul support to the small cells without requiring
additional spectrum and hardware for a separate backhaul system.
%By employing a large excess of (typically hundreds of) low power antennas,
%the radiating power of the BS is focused into ever-smaller MT spots.
The ``localized'' interference pattern of large-scale MIMO has to be considered in
cell association and wireless backhaul resource allocation design of future cellular systems.

%%%%%%%%%%%%%%%%%%%%%%%%%%%%%%%%%%%%%%%%%%%%%%%%%%%%%%%%%%%%%%%%%%%%%%%%%%%%%%%%%%%%%%%%%%%%%%%%%%
\section{System Model}\label{three}
\subsection{Large-scale MIMO Macro Base Station and Small Cells Using Wireless Backhaul}\label{3a}
%%%%%%%%%%%%%%%%%%%%%%%%%%%%%%%%%%%%%%%%%%%%%%%%%%%%%%%%%%%%%%%%%%%%%%%%%%%%%%%%%%%%%%%%%%%%%%%%%%%%%%%%%%%%%%%%
\begin{figure}
\begin{center}
\includegraphics[width=3in, draft=false]{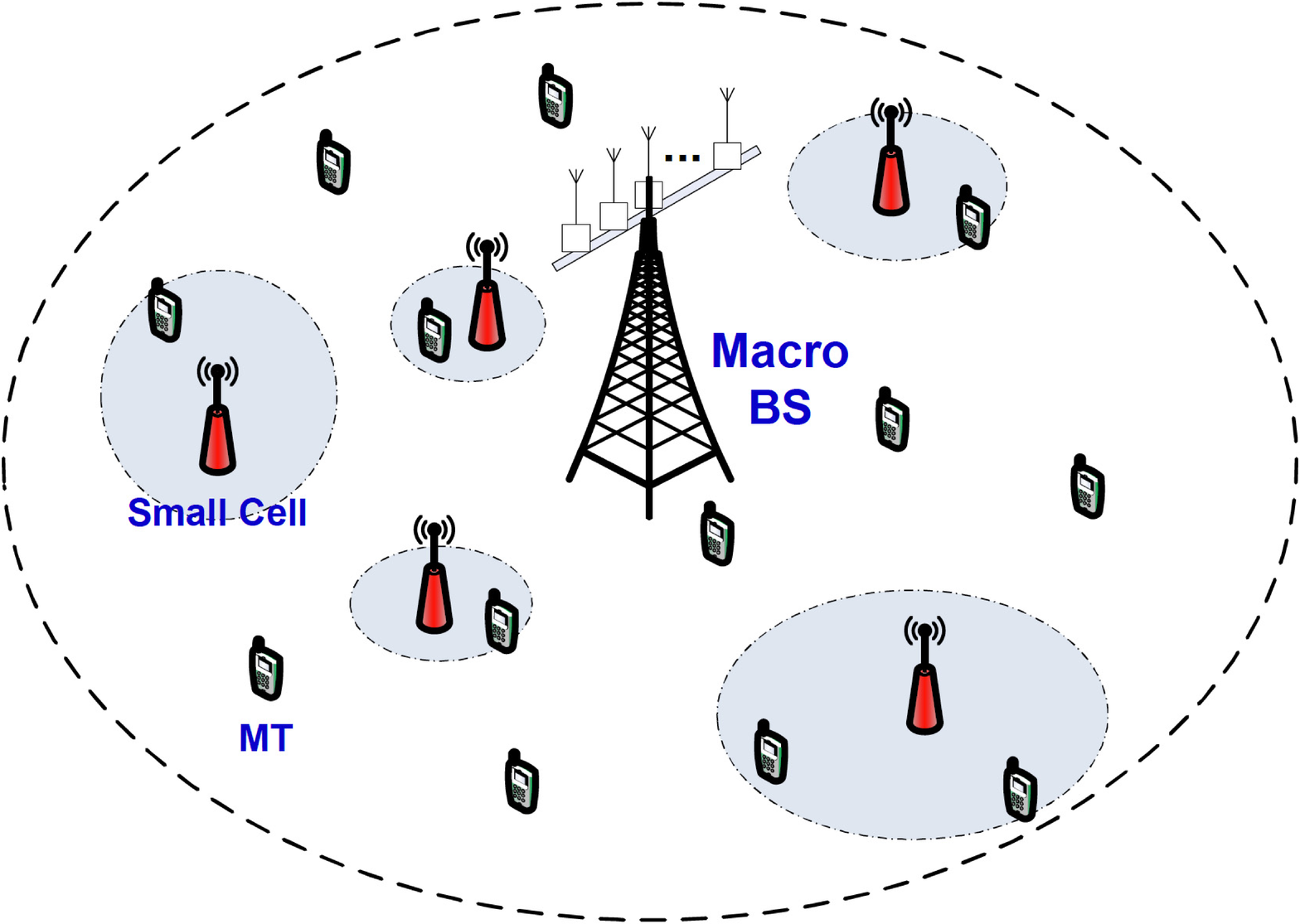}
\caption{A two-tier HetNet with a single macro cell and several small cells as well as mobile users inside the macro cell range.}
\label{HetNet}
\end{center}
\vskip -6pt
\end{figure}
%%%%%%%%%%%%%%%%%%%%%%%%%%%%%%%%%%%%%%%%%%%%%%%%%%%%%%%%%%%%%%%%%%%%%%%%%%%%%%%%%%%%%%%%%%%%%%%%%%%%%%%%%%%%%%%%
Consider a HetNet as shown in Fig. \ref{HetNet} with a single macro BS
equipped with a large array of $N_{T}$ antennas, $N_{S}$  small cells
deployed within the macro cell range, and $N_{U}$ randomly located mobile terminals.
For convenience, we denote the set of MTs by $\mathcal{U}$ and the set of small cells by $\mathcal{S}$.
$\mathcal{S}_{0}\equiv \mathcal{S}\cup\{0\}$ is the set of all BSs,
where the index $0$ is introduced for the macro BS.
As in \cite{caire}, we consider a large-scale antenna array at the macro BS having
a beamforming group size $N_{g}\ll N_{T}$.
%Without loss of generality,
We further assume that the beamforming group size $N_{g}$ is greater than $N_{S}$.
On the other hand, the number of MTs $N_{U}$ within the range of the macro BS  is much greater than $N_{g}$.
Resource sharing must be used when a large number of MTs are associated with the large-scale MIMO BS.
These assumptions render the wireless channels between the macro BS and all the other nodes orthogonal.
When all the small cells use their wireless links to the macro BS for backhauling,
the cell association scheme %is not only a means for load balancing and improving spectral efficiency.
must also take into consideration the joint resource allocation for user information and backhaul transmissions.
With orthogonality of the wireless channels to and from the large-scale antenna array at the macro BS,
high spectral efficiency is achieved with spatial-division multiplexing (SDM)
and spatial-division multiple access (SDMA).

In order to facilitate the uplink-pilot-based channel estimation of small-scale fading for
the large-scale MIMO structure at the macro BS, the two-tier HetNet in Fig. \ref{HetNet} must operate
in a TDD mode such that channel reciprocity is guaranteed \cite{massive} \cite{scaleupmimo}.
The small cells use the wireless links to the macro BS for backhauling.
Each mobile terminal selects one cell, either the macro or a small cell, to associate with.
The objective of downlink cell association is to maximize sum of the logarithmic rates of the MTs
such that a good balance between throughput maximization and fairness is achieved \cite{walrand}.

In this work, two in-band wireless backhauling strategies,
namely, unified WBBA (u-WBBA) and per-small-cell WBBA (p-WBBA), are investigated.
The former introduces a unified bandwidth allocation factor $\beta\in[0,1]$, which is
the fraction of bandwidth allocated for wireless backhauling for all the small cells within a macro cell range.
The p-WBBA scheme, on the other hand, determines $\beta_{j}\in[0,1]$, the fraction of bandwidth
allocated for wireless backhauling, for each small cell $j$ in $\mathcal{S}$.
In Sections \ref{three} and \ref{four}, we investigate joint CA-WBBA in the two-tier HetNet
with these two in-band wireless backhauling strategies.

\subsection{Wireless Channel Model}\label{3b}

With centralized massive antenna array implemented at the macro BS,
the large-scale fading is constant during one cell association period
and across a wide spectrum over all antenna elements.
Conversely, the small-scale fading
fluctuates fast enough relative to
the cell association period so that it is averaged out in the channel measurements.
We normalize the small-scale fading power and consider the average channel power gain of
the macro-BS-to-MT channels, denoted by $\bar{H}_{k}$.
%in the joint downlink CA-WBBA problem.
The channel model therefore only reflects large-scale fading, which is primarily determined by
the locations of the communicating parties and the geographical characteristics of the radio propagation paths.
We consider the non-line-of-sight (NLOS) path-loss model for urban macro BSs
suggested in 3GPP TR 36.814 V9.0.0 and log-normal shadowing \cite{3gpp_v9}.
\begin{table}
\renewcommand{\arraystretch}{0.9}
\caption{Parameter values in the radio propagation path-loss model}
\label{value} \centering
{\small
\begin{tabular}{l|l}
\hline
\bf{Parameter}     & \bf{Value} \\
\hline
\hline
Carrier frequency         & 3.5 GHz \\
System bandwidth          & 5 MHz \\
Macro BS height           & 25 m \\
Small cell BS height      & 3 m \\
Average building height   & 20 m \\
Road width                & 8 m \\
Noise power spectral density   & -174 dBm/Hz \\
%Noise power $N_{0}$            & -107 dBm \\
dB-spread of the log-normal shadowing  &  \\
\qquad  for macro BS $\sigma_{BS}$     & 6 dB \\
\qquad  for small cells $\sigma_{SC}$  & 4 dB \\
\hline
\end{tabular}
}
\end{table}

By using the model parameter values given in Table \ref{value},
the large-scale channel gain $\bar{H}_{k}$ in dB is specified as
\begin{equation}\label{Hk}
\bar{H}_{k} = 27.3 + 3.91\times10\log_{10}d_{k} + Z(\sigma^{2}_{BS}),
\end{equation}
where $d_{k}$ is the link distance in meters from the macro BS to the $k^{\rm th}$ MT,
$Z(\sigma^{2}_{BS})$ is the log-normal shadowing term in dB,
which is a zero-mean Gaussian random variable with variance $\sigma^{2}_{BS}$.
Similarly, we use $\bar{G}_{j}$ and $\bar{L}_{j,k}$ to characterize
the wireless channel between the macro BS and the $j^{\rm{th}}$ small cell
and that between the $j^{\rm{th}}$ small cell and the $k^{\rm{th}}$ MT, respectively.
$\bar{G}_{j}$ and $\bar{L}_{j,k}$ also account for large-scale fading only.
The downlink wireless backhaul path-loss $\bar{G}_{j}$ for small cell $j$ in dB is
\begin{equation}\label{Gj}
\bar{G}_{j} = 24.6 + 3.91\times10\log_{10}d_{j} + Z(\sigma^{2}_{BS}),
\end{equation}
where $d_{j}$ represents the link distance in meters from the macro BS to the $j^{\rm th}$ small cell.
A NLOS urban micro BS propagation model suggested in \cite{3gpp_v9} is adopted for downlink in small cells,
which gives
\begin{equation}\label{Ljk}
\bar{L}_{j,k} = 36.8 + 3.67\times10\log_{10}d_{j,k} + Z(\sigma^{2}_{SC}),
\end{equation}
where $d_{j,k}$ is the link distance in meters from small cell $j$ to the $k^{\rm th}$ MT.
As in (\ref{Hk}) and (\ref{Gj}), $Z(\sigma^{2}_{SC})$ is the log-normal shadowing term in dB,
which is zero-mean Gaussian with variance $\sigma^{2}_{SC}$.
%According to \cite{3gpp_v9}, here we use dB-spread values $\sigma_{BS}=6$ dB and $\sigma_{SC}=4$ dB.

\subsection{Interference Management Considerations for Wireless Backhauling of Small Cells}\label{3c}

\begin{figure}
\begin{center}
\includegraphics[width=4in, draft=false]{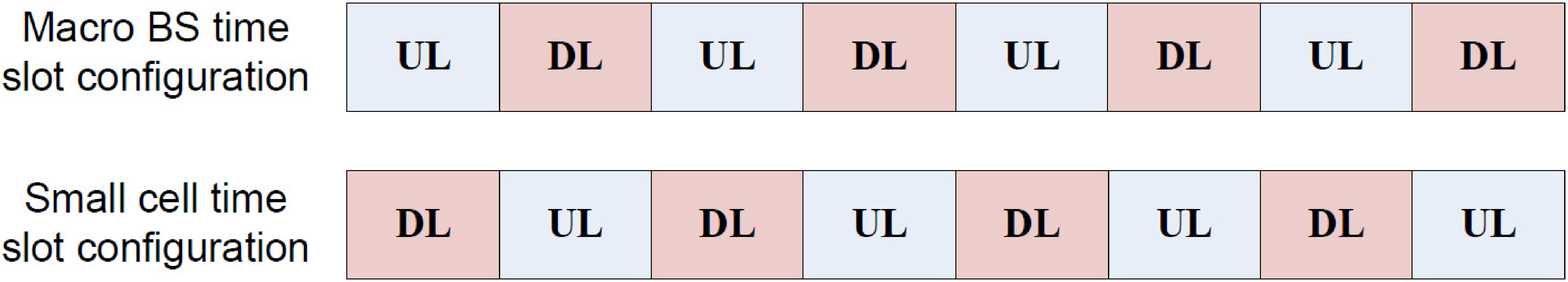}
\caption{An example of the reverse-TDD time slot configuration for macro BS and small cells in a two-tier HetNet.}
\label{rvTDD}
\end{center}
%\vskip -6pt
\end{figure}

As discussed in Section \ref{2a}, the TDD mode for duplexing is employed by the two-tier HetNet
to facilitate the massive MIMO feature of the macro BS.
Several TDD schemes for HetNets with large-scale antenna array macro BSs were studied in \cite{reversetdd}.
The co-channel reverse-TDD scheme, which uses reversed uplink/downlink (UL/DL) time slot configurations for small cells and macro BSs,
has a preferred inter-cell interference pattern and thus is a suitable model for this work.
A simple example of the time slot configurations for macro and small cells operating
in the reverse-TDD mode is shown in Fig. \ref{rvTDD}.

%%%%%%%%%%%%%%%%%%%%%%%%%%%%%%%%%%%%%%%%%%%%%%%%%%%%%%%%%%%%%%%%%%%%%%%%%%%%%%%%%%%%%%%%%%%%%%%%%%%%%%%%%%%%%%%%
%%%%%%%%%%%%%%%%%%%%%%%%%%%%%%%%%%%%%%%%%%%%%%%%%%%%%%%%%%%%%%%%%%%%%%%%%%%%%%%%%%%%%%%%%%%%%%%%%%%%%%%%%%%%%%%%

The reverse-TDD scheme is a natural  choice because of the use of in-band wireless backhauling of small cells.
For convenience, we call the small cell's functionality for communicating with its associated MTs as the {\bf service functionality},
and the functionality for in-band wireless backhaul communications as the {\bf backhaul functionality}.
From the wireless backhaul's standpoint, the small cells act as mobile users associated with the macro BS.
%when they use their wireless links to the macro base station for backhaul.
During a downlink time slot designated by the macro BS, the small cells must be in receiving mode for the backhauling functionality.
When a small cell is in receiving mode, for backhauling,
it should at the same time receive transmissions from the MTs, on frequency channels orthogonal to those used for wireless backhauling.
Otherwise, significant self-interference will occur at the small cell.
In this case, we see that this time slot has to be an uplink time slot of the small cell as a serving base station.
Similarly, when a small cell is in transmitting mode, it simultaneously transmits to its associated MTs
and the wireless backhaul hub (the macro BS) on orthogonal frequency channels.
The time slot is therefore a downlink time slot to its own associated MTs,
while it is an uplink time slot for the macro BS.
Therefore, in a small cell using wireless backhauling,
we have to reverse the DL/UL time slot configurations designated by the macro BS.
In this work, we consider that all the time slots are equally allocated for uplink and downlink.
All the small cells use the same DL/UL time slot configuration,
which is the reversed version of that designated by the macro BS.

In cellular systems operating under TDD mode, severe co-channel interference may occur at cell edges
if different time slot configurations are employed by adjacent cells.
This can be shown by the following example,
which is similar to the cross time slot scenario in TD-SCDMA networks \cite{tdscdma}.
%%%%%%%%%%%%%%%%%%%%%%%%%%%%%%%%%%%%%%%%%%%%%%%%%%%%%%%%%%%%%%%%%%%%%%%%%%%%%%%%%%%%%%%%%%%%%%%%%%%%%%%%%%%%%%%%
\begin{figure}\label{crossts}
\begin{center}
\includegraphics[width=3.5in, draft=false]{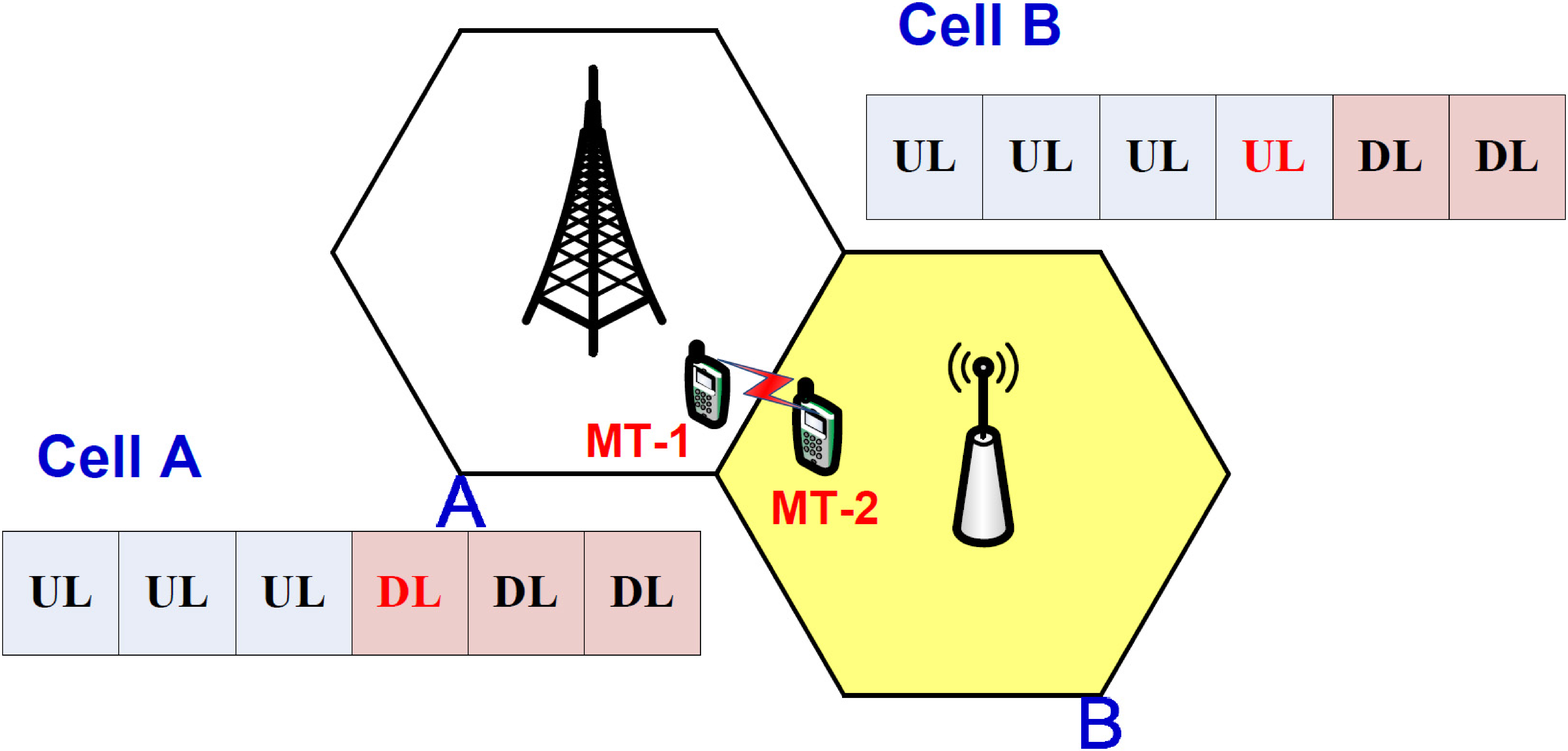}
\caption{Cell edge inter-cell interference between adjacent cells in a TDD system when
conflicting UL/DL time slot configuration occurs.}
\end{center}
\vskip -12pt
\end{figure}
%%%%%%%%%%%%%%%%%%%%%%%%%%%%%%%%%%%%%%%%%%%%%%%%%%%%%%%%%%%%%%%%%%%%%%%%%%%%%%%%%%%%%%%%%%%%%%%%%%%%%%%%%%%%%%%%
With the DL/UL assignments as shown in Fig. 3 for adjacent cells A and B,
it is a DL time slot in Cell A and a UL time slot in Cell B in the fourth time slot.
If there are cell edge users MT-1 (associated with Cell A) and MT-2 (associated with Cell B),
which are close to each other, MT-2's uplink transmission will introduce significant co-channel interference
to MT-1's downlink reception in time slot number four.
In our problem, where the small cells use cellular links to the macro BS for backhauling,
the worst-case (adjacent cell) co-channel inter-cell interference occurs between a small cell and its associated MTs.
That is, the small cell's uplink backhaul transmission (to the macro BS) will interfere with
its downlink communication with the associated MTs if they are assigned the same frequency band.
Similarly, uplink transmissions of a small cell's associated MTs will interfere with
the macrocell-to-smallcell backhaul link at the small cell receiver.
Because the two interfering communication links share a common node, i.e., the small cell BS,
universal frequency reuse is out of the question.
Soft frequency reuse (SFR) \cite{sfr} is a good compromise between spectral efficiency
and interference control in such circumstances.
Specifically, in this work, we consider that the bandwidth allocated for wireless backhauling is not reused for user transmissions,
either globally (in the u-WBBA case) or locally (in the p-WBBA case).

Assume that the total frequency band can be divided into orthogonal sub-frequency-bands.
In a multi-user MIMO system operating in the ``large-scale MIMO'' regime, i.e., $N_{T}$
is much greater than the number of simultaneous associated terminals, SDM within a macro cell range is considered.
A beamforming group size $N_{g}$ is available for each sub-frequency-band.
Therefore, when the number of associated terminals for the large-scale MIMO macro BS is large,
joint frequency-division multiple access (FDMA) and SDMA is employed.
On the other hand, in a small cell,
bandwidth allocation must be performed between communication with its associated MTs and the wireless backhaul
to avoid severe co-channel interference between the two functionalities.

%This reverse-TDD-SFR framework proposed for interference management also helps to
%balance the load of small cells and their capability to serve mobile terminals.
%Bandwidth allocation in small cells is conducted periodically in the same time scale as cell association.

%%%%%%%%%%%%%%%%%%%%%%%%%%%%%%%%%%%%%%%%%%%%%%%%%%%%%%%%%%%%%%%%%%%%%%%%%%%%%%%%%%%%%%%%%%%%%%%%%%
\section{Joint Cell Association and Unified Bandwidth Allocation for Wireless Backhauling of Small Cells}\label{four}

\subsection{Problem Formulation for Joint Optimization}\label{4a}

We consider a ``large-scale MIMO'' regime at the macro BS for its downlink transmissions.
The two-tier HetNet in Fig. \ref{HetNet} operates under reverse-TDD mode.
A binary indicator variable $x_{j,k}\in\{0,1\}$ is introduced for cell association status
of mobile terminal $k\in\mathcal{U}$ with the $j^{\rm{th}}$ cell ($j\in\mathcal{S}_{0}$),
where 1 indicates association.
It is straightforward that $x_{j,k}$ must satisfy
\begin{equation}\label{sumx}
\sum_{j\in\mathcal{S}_{0}}x_{j,k}= 1,\ \forall k\in\mathcal{U}.
\end{equation}

According to the proposed reverse-TDD-SFR framework for interference management,
downlink transmissions to the MTs associated with the macro BS are only interfered by small cell users,
which is sufficiently small to be ignored.
The downlink of a small cell, on the other hand, is subject to interference from other small cells' downlink transmissions.
Therefore, we define the downlink SINR in the macro cell as $\gamma_{0,k}=\frac{P_{0}\bar{H}_{k}}{N_{0}}$,
where $P_{0}$ is the transmit power of the macro BS.
Similarly, the SINR in the wireless backhaul downlink for small cell $j$ is $\gamma_{j}=\frac{P_{0}\bar{G}_{j}}{N_{0}}$.
The received SINR of MT $k$ associated with small cell $j$ is
$\gamma_{j,k}=\frac{P_{j}\bar{L}_{j,k}}{N_{0}+\sum_{l\in\mathcal{S},l\neq j}P_{l}\bar{L}_{l,k}}$,
where $P_{j}$ ($P_{l}$) denotes the transmit power of small cell $j$ ($l$).

When multiple MTs are associated with a single-antenna small cell $j\in\mathcal{S}$,
equal resource allocation among these MTs is optimal \cite{ye}.
For the u-WBBA scenario,
the fraction of bandwidth allocated for wireless backhauling is $\beta\in[0,1]$.
The remaining bandwidth is used for user data transmission.
The throughput of the $j^{\rm{th}}$ small cell is therefore given by
\begin{equation}\label{Rj}
R_{j}^{U} = \sum_{k\in\mathcal{U}}R_{j,k}^{U}=\frac{1-\beta}{\sum_{k\in\mathcal{U}}x_{j,k}}
\sum_{k\in\mathcal{U}}x_{j,k}\log_{2}(1+\gamma_{j,k}),
\end{equation}
where
%$P_{j}$ is the average transmit SNR of the $j^{\rm{th}}$ small cell.
the superscript ``U'' indicates the unified WBBA scenario.
Because of the wireless backhaul constraint, the throughput $R_{j}^{U}$ given by (\ref{Rj})
must not exceed the achievable rate of the wireless communication link
between the macro BS and the $j^{\rm{th}}$ small cell, which will be analyzed in the following.

In the ``large-scale MIMO'' regime, the macro BS has an antenna array size $N_{T}$
much greater than the number of simultaneously served nodes (including both MTs and the small cells).
In this work, we use the approximate model in \cite{caire} and
assume $N_{T}\gg N_{g}$ and $N_{U}>N_{g}$.
A large number of MTs associated with the macro BS is served by resource sharing, e.g., time sharing.
When a mobile terminal $k$ is associated with the macro BS,
its achievable rate $R_{0,k}$ in the ``large-scale MIMO'' regime is determined by both
the channel coefficient $\bar{H}_{k}$,
and the number of MTs associated with the massive MIMO macro BS \cite{caire}.
Given the unified small cell bandwidth allocation factor $\beta$,
the achievable rate of the $k^{\rm{th}}$ MT associated with the macro cell is \cite{caire}
\begin{equation}\label{R0k}
R_{0,k}^{U} = x_{0,k}\frac{(1-\beta)N_{g}}{\sum_{k\in\mathcal{U}}x_{0,k}}
\log_{2}\left(1+\frac{N_{T}-N_{g}+1}{N_{g}}\gamma_{0,k}\right),
\end{equation}
where %$P_{0}$ denotes the transmit SNR of the macro BS, and
equal resource sharing among MTs associated with the macro BS is assumed.
Similarly, the capacity of the wireless backhaul downlink for small cell $j$ is given by
\begin{equation}\label{Cj}
C_{j}^{U} = \beta\log_{2}\left(1+\frac{N_{T}-N_{g}+1}{N_{g}}\gamma_{j}\right).
\end{equation}
The downlink wireless backhaul constraint requires $R_{j}^{U}\leq C_{j}^{U}$ such that
all the downlink traffic of the associated MTs of small cell $j$ in the current cell association period
can be accommodated by its wireless backhaul.

Note that as illustrated in Section \ref{3c}, in order to facilitate the wireless backhaul design,
a reverse-TDD scheme for duplexing is introduced for the two-tier HetNet.
As a consequence, consider a time slot configuration as in Fig. \ref{rvTDD},
the information a small cell transmits to its associated MTs in a DL time slot of the small cell
is that it received through the wireless backhaul from the macro BS in the previous time slot,
which is a DL time slot of the macro BS.
Because the radio propagation path-loss and large-scale fading are constant over several time slots,
the inequality $R_{j}^{U}\leq C_{j}^{U}$ reflects the overall requirement on the downlink wireless backhaul
for the entire cell association period containing a number of time slots.

For convenience, we denote
\[
r_{j,k}=\log_{2}\left(1+\gamma_{j,k}\right), \forall(j,k)\in\mathcal{S}\times\mathcal{U},
\]
\[
r_{0,k}=N_{g}\log_{2}\left(1+\frac{N_{T}-N_{g}+1}{N_{g}}\gamma_{0,k}\right), \forall k\in\mathcal{U},
\]
and
\[
c_{j}=\log_{2}\left(1+\frac{N_{T}-N_{g}+1}{N_{g}}\gamma_{j}\right), \forall j\in\mathcal{S},
\]
which can be considered as base-line rate expressions of the downlink channels
that do not change with $x_{j,k}$s or $\beta$.
The optimization objective, which is selected as the network's sum log-rate, as discussed in Section \ref{2a},
to achieve proportional fairness, is given by %(\ref{logthroughput}).
\begin{equation}\label{logthroughput}
%\begin{aligned}
\bar{R}^{U} = %\sum_{k\in\mathcal{U}}x_{0,k}\log\left[\frac{(1-\beta)r_{0,k}}{\sum_{k\in\mathcal{U}}x_{0,k}}\right]\\
\sum_{j\in\mathcal{S}_{0}}\sum_{k\in\mathcal{U}}x_{j,k}
\log\left[\frac{(1-\beta)r_{j,k}}{\sum_{k\in\mathcal{U}}x_{j,k}}\right]
%&= \sum_{k\in\mathcal{U}}x_{0,k}\bar{R}_{0,k} + \sum_{j\in\mathcal{S}}\sum_{k\in\mathcal{U}}x_{j,k}\bar{R}_{j,k}.
%&= \bar{R}_{0}^{U} +
=\sum_{j\in\mathcal{S}_{0}}\bar{R}_{j}^{U}.
%\end{aligned}
\end{equation}

We denote $\mathbf{X}=\{x_{j,k};\ j\in\mathcal{S}_{0}, k\in\mathcal{U}\}$ and
have the following optimization problem $\mathbf{P1}$ for joint downlink CA-WBBA with unified
wireless backhaul bandwidth allocation factor,
which is a mixed-integer nonlinear programming (MINLP) problem.
\begin{eqnarray}\label{opt1}
%\begin{array}{ll}
\mathbf{P1:}\ \ \underset{\beta,\mathbf{X}}{\mbox{maximize}}
                                &&\bar{R}^{U}(\beta,\mathbf{X}) \nonumber \\
\label{constraint1}
\mbox{subject to}               &&x_{j,k}\in\{0,1\},\ \forall(j,k)\in\mathcal{S}_{0}\times\mathcal{U};\\
\label{constraint2}
\                               &&\sum_{j\in\mathcal{S}_{0}}x_{j,k}=1,\ \forall k\in\mathcal{U};\\
\label{constraint3}
\                               &&0\leq \beta\leq 1;\\
\label{constraint4}
\                               &&R_{j}^{U}\leq C_{j}^{U},\ \forall j\in\mathcal{S}.
%\end{array}
\end{eqnarray}
For the u-WBBA scenario, the parameters $R_{j}^{U}$, $C_{j}^{U}$, and $\bar{R}^{U}$ in $\mathbf{P1}$
are specified as in (\ref{Rj})--(\ref{logthroughput}).

It can be shown that the objective function in (\ref{logthroughput}) is concave,
and the continuous variable $\beta$ and the binary variables $x_{j,k}$s are separable in (\ref{logthroughput}).
However,
the constraint in (\ref{constraint4}) is a nonlinear non-convex coupling constraint.
By fixing the bandwidth allocation factor $\beta$, the constraint in (\ref{constraint4}) becomes convex constraint on $x_{j,k}$s.
Therefore, we will relax the cell association variables $x_{j,k}$s
and consider a decomposition approach to solve the joint CA-WBBA problem with one unified WBBA factor $\beta$
for all the small cells.

\subsection{Relaxed Optimization and Hierarchical Decomposition}\label{4b}
We relax the binary cell association indicators $x_{j,k}$s in problem $\mathbf{P1}$,
which gives
\begin{equation}
0\leq x_{j,k}\leq 1,\ \forall(j,k)\in\mathcal{S}_{0}\times\mathcal{U}.
\end{equation}
Fractional cell association indicators can be interpreted as
partial association with different cells in a cell association period.
A hierarchical decomposition approach \cite{palomar}, which contains an upper level primal decomposition
and a lower level dual composition, is considered to solve the relaxed optimization problem.
By first conducting an {\bf upper level primal decomposition} for the problem with relaxed association variables,
we obtain an inner problem for cell association,
and an outer problem for unified wireless backhaul bandwidth allocation.
%in each iteration of the algorithm execution.

Given a $\beta$ value for unified WBBA, the optimization algorithm begins with
the inner problem $\mathbf{P1.1}$ formulated as
\begin{eqnarray}\label{opt1.1}
\mathbf{P1.1:}\ \ \underset{\mathbf{X}}{\mbox{maximize}}
                                %&&\sum_{j\in\mathcal{S}_{0}}\sum_{k\in\mathcal{U}}x_{j,k}\log\frac{r_{j,k}}{\sum_{k\in\mathcal{U}}x_{j,k}}\nonumber \\
                                &&\sum_{j\in\mathcal{S}_{0}}\bar{R}_{j}^{U'}(\mathbf{X})\nonumber \\
\label{constraint1.1.1}
\mbox{subject to}               &&x_{j,k}\geq 0,\ \forall(j,k)\in\mathcal{S}_{0}\times\mathcal{U};\\
\label{constraint1.1.2}
\                               &&\sum_{j\in\mathcal{S}_{0}}x_{j,k}=1,\ \forall k\in\mathcal{U},\\
\label{constraint1.1.3}
\                               &&R_{j}^{U}(\mathbf{X};\beta)-C_{j}^{U}(\mathbf{X};\beta)\leq 0, \forall j\in\mathcal{S},
\end{eqnarray}
where $\bar{R}_{j}^{U'}$s, defined by $\bar{R}_{j}^{U'}=\sum_{k\in\mathcal{U}}x_{j,k}\log\frac{r_{j,k}}{\sum_{k\in\mathcal{U}}x_{j,k}}$,
are concave functions of $x_{j,k}$s.
%As shown in Appendix \ref{Appendix-A},
%$\bar{R}_{j}^{U'}$ is a concave function for all $j\in\mathcal{S}_{0}$,
Because all constraints in $\mathbf{P1.1}$ are linear,
the inner problem $\mathbf{P1.1}$ is a convex optimization problem.
%which can be effectively solved by either centralized optimization
%or distributed algorithms via dual decomposition \cite{ye}.

Once the optimal solution $\mathbf{X}^{\star}=\{x_{j,k}^{\star}; (j,k)\in\mathcal{S}_{0}\times\mathcal{U}\}$
is obtained for the convex inner problem $\mathbf{P1.1}$ parameterized by $\beta$,
it can be used in the following outer problem $\mathbf{P1.2}$ for the unified wireless backhaul
bandwidth allocation:
\begin{eqnarray}\label{opt1.2}
\mathbf{P1.2:}\ \ \underset{\beta}{\mbox{maximize}}
                                &&N_{U}\log(1-\beta) %+ \bar{R}_{0}^{U'}(\mathbf{X}^{\star})
                                + \sum_{j\in\mathcal{S}_{0}}\bar{R}_{j}^{U'}(\mathbf{X}^{\star})\nonumber \\
\label{constraint1.2.1}
\mbox{subject to}               &&0\leq \beta\leq 1;\\
\label{constraint1.2.2}
\                               &&R_{j}^{U}(\beta;\mathbf{X}^{\star})-C_{j}^{U}(\beta;\mathbf{X}^{\star})\leq 0,j\in\mathcal{S},
\end{eqnarray}
where $\bar{R}_{j}^{U'}(\mathbf{X}^{\star})$ is the function value of $\bar{R}_{j}^{U'}$ evaluated at
$\mathbf{X}^{\star}$.
In order to obtain the solution to the master problem $\mathbf{P1}$ for joint CA-WBBA with u-WBBA,
the two sub-problems $\mathbf{P1.1}$ and $\mathbf{P1.2}$ are solved iteratively until convergence.
We call the iterations for solving $\mathbf{P1.1}$ and $\mathbf{P1.2}$ for the primal decomposition
the {\bf outer iterations}, which are different from the {\bf inner iterations} of the lower level
dual decomposition algorithm to be discussed in more detail in Section \ref{3c}.

Note that $\bar{R}_{j}^{U'}$s are constants in the outer problem $\mathbf{P1.2}$.
Maximizing the objective function of $\mathbf{P1.2}$ with respect to $\beta$ is equivalent to maximizing $\log(1-\beta)$ only,
which is a concave and monotonically decreasing function of $\beta$.
Problem $\mathbf{P1.2}$ then reduces to a feasibility problem whose solution is the smallest feasible value of $\beta$ given the constraints
(\ref{constraint1.2.1}) and (\ref{constraint1.2.2}).
By invoking the wireless backhaul constraint in (\ref{constraint1.2.2}),
the u-WBBA factor $\beta$ must satisfy
\begin{equation}\label{betau}
\beta \geq \max\left\{\frac{\sum_{k\in\mathcal{U}}x_{j,k}^{\star}r_{j,k}}{\sum_{k\in\mathcal{U}}x_{j,k}^{\star}(r_{j,k}+c_{j})},\ \forall j\in\mathcal{S}\right\}.
\end{equation}
The solution to the outer problem $\mathbf{P1.2}$, i.e., the smallest feasible $\beta$ which satisfies the wireless backhaul constraint,
is therefore given by (\ref{betau}) at equality.
Since the complexity of evaluating (\ref{betau}) is low,
and the number of small cells within a macro cell range is limited by the massive MIMO beamforming group size $N_{g}$,
which is typically not a very large number,
globally optimal $\beta$ for the outer problem $\mathbf{P1.2}$ can be easily identified.
The complexity of solving the master problem $\mathbf{P1}$ with the primal decomposition method
lies mainly in solving the convex inner problem $\mathbf{P1.1}$.

In order to solve the convex problem $\mathbf{P1.1}$ for cell association
in a distributed manner such that the need for complex inter-tier coordination is avoided,
we consider a distributed algorithm considered in \cite{ye},
in which a {\bf lower level dual decomposition} will be conducted for the inner problem $\mathbf{P1.1}$.
Auxiliary variables $K_{j}=\sum_{k\in\mathcal{U}}x_{j,k},\ \forall j\in\mathcal{S}_{0}$ are introduced
so that the inner problem $\mathbf{P1.1}$ is equivalently given as
\begin{eqnarray}\label{opt1.1r}
\mathbf{P1.1r:}\ \ \underset{\mathbf{X},\{K_{j}, j\in\mathcal{S}_{0}\}}{\mbox{maximize}}
\label{obj1.1r}
                                &&\sum_{j\in\mathcal{S}_{0}}\sum_{k\in\mathcal{U}}x_{j,k}\log(r_{j,k})
 - \sum_{j\in\mathcal{S}_{0}}K_{j}\log(K_{j}) \\
\label{constraint1.1.1r}
\mbox{subject to}               &&x_{j,k}\geq 0,\ \forall(j,k)\in\mathcal{S}_{0}\times\mathcal{U};\nonumber \\
\label{constraint1.1.2r}
\                               &&\sum_{j\in\mathcal{S}_{0}}x_{j,k}=1,\ \forall k\in\mathcal{U};\nonumber \\
\label{constraint1.1.3r}
                                &&K_{j}=\sum_{k\in\mathcal{U}}x_{j,k},\ \forall j\in\mathcal{S}_{0};\\
\label{constraint1.1.4r}
                                &&0\leq K_{j}\leq N_{U},\ \forall j\in\mathcal{S}_{0};\\
\label{constraint1.1.5r}
\                               &&R_{j}^{U}(\mathbf{X};\beta)-C_{j}^{U}(\mathbf{X};\beta)\leq 0, j\in\mathcal{S}.
\end{eqnarray}
By jointly considering the objective function in (\ref{obj1.1r}) and the coupling constraints in (\ref{constraint1.1.3r})
and (\ref{constraint1.1.5r}), we obtain the Lagrangian in (\ref{lagrU}) associated with problem $\mathbf{P1.1r}$,
in which we have $\mu=[\mu_{0},\mu_{1},\ldots,\mu_{N_{S}}]^{T}$ and $\nu=[0,\nu_{1},\ldots,\nu_{N_{S}}]^{T}$.

\begin{equation}\label{lagrU}
\begin{aligned}
&\quad\ \mathcal{L}^{U}(\mathbf{X},\mu,\nu)\\
&= \sum_{j\in\mathcal{S}_{0}}\sum_{k\in\mathcal{U}}x_{j,k}\log(r_{j,k})
- \sum_{j\in\mathcal{S}_{0}}K_{j}\log(K_{j})
+ \sum_{j\in\mathcal{S}_{0}}\mu_{j}(K_{j}-\sum_{k\in\mathcal{U}}x_{j,k})
+ \sum_{j\in\mathcal{S}_{0}}\nu_{j}\left[\beta c_{j}K_{j}-(1-\beta)\sum_{k\in\mathcal{U}}x_{j,k}r_{j,k}\right]\\
&= \sum_{k\in\mathcal{U}}\sum_{j\in\mathcal{S}_{0}}x_{j,k}\left[\log(r_{j,k})-\mu_{j}-\nu_{j}(1-\beta)r_{j,k}\right]
+ \sum_{j\in\mathcal{S}_{0}}K_{j}\left[\mu_{j}-\log(K_{j})+\nu_{j}\beta c_{j}\right].
\end{aligned}
\end{equation}
The corresponding Lagrangian dual function is therefore given as
\begin{equation}\label{gmunu}
g(\mu,\nu) = \sum_{k\in\mathcal{U}}g_{k}(\mu,\nu) + g_{K}(\mu,\nu),
\end{equation}
where we define
\begin{subequations}
\begin{eqnarray}
\label{gkmu}
g_{k}(\mu,\nu) = \underset{x_{j,k},j\in\mathcal{S}_{0}}{\sup} &&\sum_{j\in\mathcal{S}_{0}}x_{j,k}\left(
%\begin{aligned}
\log(r_{j,k})-\mu_{j}
-\nu_{j}(1-\beta)r_{j,k}
%\end{aligned}
\right)\\
%\right].\nonumber\\
%&&\qquad\qquad \left.+\nu_{j}(1-\beta)r_{j,k}\right]\\
\mbox{subject to}  &&x_{j,k}\geq 0,\ \forall j\in\mathcal{S}_{0}; \nonumber\\
\                  &&\sum_{j\in\mathcal{S}_{0}}x_{j,k}=1. \nonumber
\end{eqnarray}
\begin{eqnarray}
\label{gKKmu}
g_{K}(\mu,\nu) = \underset{K_{j},j\in\mathcal{S}_{0}}{\sup}  &&\sum_{j\in\mathcal{S}_{0}}K_{j}\left(
%\begin{aligned}
\mu_{j}-\log(K_{j})
+\nu_{j}\beta c_{j}
%\end{aligned}
\right)\\
%\right.\nonumber\\
%&&\qquad\qquad \left.-\nu_{j}\beta c_{j}\right]\\
\mbox{subject to}  &&K_{j}\leq N_{U},\ \forall j\in\mathcal{S}_{0}. \nonumber
\end{eqnarray}
\end{subequations}
The following dual problem for cell association minimizes the Lagrangian dual function $g(\mu,\nu)$ with respect to $\mu$ and $\nu$:
\begin{eqnarray}\label{optD1}
\mathbf{D1.1:}\ \ \underset{\mu,\nu}{\mbox{minimize}}
                   &&\sum_{k\in\mathcal{U}}g_{k}(\mu,\nu) + g_{K}(\mu,\nu).
\end{eqnarray}
Having only linear constraints in the convex problem $\mathbf{P1.1r}$, the Slater's condition
is straightforward and thus strong duality holds \cite{boyd}.
The primal problem $\mathbf{P1.1r}$ can therefore be equivalently solved by solving the dual problem $\mathbf{D1.1}$.

\subsection{Distributed Joint CA-WBBA Method With Unified WBBA}\label{4c}

Based on the hierarchical decomposition approach investigated in Section \ref{4b},
we next present our method for distributed joint CA-WBBA in a two-tier HetNet
employing the u-WBBA strategy.

As illustrated in Section \ref{4b}, solving the inner problem $\mathbf{P1.1}$ for cell association is
equivalent to solving the Lagrangian dual problem $\mathbf{D1.1}$.
According to (\ref{gkmu}), given $\mu_{j}^{(t)}$s and $\nu_{j}^{(t)}$s obtained from the $t^{\rm th}$ inner iteration,
the sub-problem $g_{k}(\mu^{(t)},\nu^{(t)})$ implies that each user $k\in\mathcal{U}$ simply chooses the BS
that offers the highest revenue $Q_{j,k}^{(t)}=\log(r_{j,k})-\mu_{j}^{(t)}-\nu_{j}^{(t)}(1-\beta)r_{j,k}$.
This mechanism for updating $x_{j,k}$s is expressed as follows
\begin{equation}\label{xtp1}
x_{j,k}^{(t+1)} = \left\{
\begin{array}{ll}
1,     &j=j_{k}^{(t)}\\
0,     &j\neq j_{k}^{(t)}
\end{array}
\right.,\ \forall k\in\mathcal{U},
\end{equation}
where
\begin{equation}\label{jkt}
j_{k}^{(t)} = \underset{j\in\mathcal{S}_{0}}{\arg\max}
\left[\log(r_{j,k})-\mu_{j}^{(t)}-\nu_{j}^{(t)}(1-\beta)r_{j,k}\right],\ \forall k\in\mathcal{U}.
\end{equation}
Note that according to (\ref{xtp1}), the algorithm only assigns binary values to
the cell association indicator variables $x_{j,k}$s without introducing any form of relaxation.
The proposed hierarchical decomposition approach actually gives exact solution to the MINLP problem
for joint CA-WBBA problem with unified WBBA.

On the other hand, $K_{j}$s can be updated locally at each BS by
invoking the first order condition for (\ref{gKKmu}) as
\begin{equation}\label{Kjp1}
K_{j}^{(t+1)} = \min\left\{\exp(\mu_{j}^{(t)}+\nu_{j}^{(t)}\beta c_{j}-1),N_{U}\right\},\ \forall j\in\mathcal{S}_{0}.
\end{equation}
It is noted that the Lagrangian dual function $g(\mu,\nu)$ given in (\ref{gmunu})
is not differentiable on $\mu$ or $\nu$.
A sub-gradient approach is used here to update the Lagrangian multipliers in the algorithm.
Specifically, with carefully chosen step sizes $\delta_{\mu}(t)$ and $\delta_{\nu}(t)$,
which are usually quantities that diminish with the iteration count $t$,
the Lagrangian multipliers are updated as
\begin{equation}\label{mutp1}
\mu_{j}^{(t+1)} = \left[\mu_{j}^{(t)}-\delta_{\mu}(t)\left(K_{j}^{(t+1)}-\sum_{j\in\mathcal{S}_{0}}x_{j,k}^{(t+1)}\right)\right]^{+},
\end{equation}
and
\begin{equation}\label{nutp1}
%\begin{aligned}
\nu_{j}^{(t+1)} = \left[\nu_{j}^{(t)}-\delta_{\nu}(t)
\left(\beta c_{j}K_{j}^{(t+1)}-(1-\beta)\sum_{k\in\mathcal{U}}x_{j,k}^{(t+1)}r_{j,k}\right)\right]^{+},
%\end{aligned}
\end{equation}
where the operator $[\cdot]^{+}$ picks the maximum of zero and its argument.
Convergence of the sub-gradient solution to the dual problem $\mathbf{D1.1}$ is guaranteed
with appropriate step sizes $\delta_{\mu}(t)$ and $\delta_{\nu}(t)$ \cite{ye} \cite{bertsekas}.
By alternatively updating the primal and dual variables until convergence,
we obtain the solution to the inner problem $\mathbf{P1.1}$ for cell association, denoted by $\mathbf{X}^{\star}$.

The outer iterations in the master problem $\mathbf{P1}$ are conducted for problems $\mathbf{P1.1}$ and $\mathbf{P1.2}$,
which are obtained from the upper level primal decomposition.
Specifically, the cell association solution obtained from the inner iterations for problem $\mathbf{P1.1}$
is fed back to the outer problem $\mathbf{P1.2}$.
Based on our analyses in Section \ref{3b}, the unified WBBA factor is updated in the outer iteration as
\begin{equation}
\beta = \max\left\{\frac{\sum_{k\in\mathcal{U}}x_{j,k}^{\star}r_{j,k}}{\sum_{k\in\mathcal{U}}x_{j,k}^{\star}(r_{j,k}+c_{j})},\ \forall j\in\mathcal{S}\right\},
\end{equation}
which is the $\beta$ value to be used for problem $\mathbf{P1.1}$ in the next outer iteration.

We summarize the distributed method for joint CA-WBBA with u-WBBA
based on our hierarchical decomposition solution in \textbf{Algorithm \ref{alg0}}.
The operations and responses of different network entities, i.e., the macro BS, small cells, and MTs,
are specified in the design.
\begin{algorithm}
\caption{Distributed method for joint CA-WBBA with unified WBBA}\label{alg0}
\begin{algorithmic}
\State {{\bf Input:}  A feasible initial value of the unified WBBA factor $\beta$.
Large-scale channel coefficients $\bar{H}_{k}$s, $\bar{G}_{j}$s, and $\bar{L}_{j,k}$s
for all $j\in\mathcal{S}$ and $k\in\mathcal{U}$.}
\State {{\bf Initialization:} Calculate $r_{j,k}$s, $r_{0,k}$s, and $c_{j}$s, for all $j\in\mathcal{S}$ and $k\in\mathcal{U}$.}
\State {{\bf Output:} Optimal cell association $\mathbf{X}^{\star}$ and WBBA factor $\beta^{\star}$.}
\State {{\bf Execution:}}
\While {Convergence not achieved (outer iteration $s$)}
\While {Convergence not achieved (inner iteration $t$)}
\State {{\bf -- MT strategy (inner iteration):}}
\For { $\forall k\in\mathcal{U}$ }
\State {Calculate $\mathbf{X}^{(t+1)}$ according to (\ref{jkt}) and (\ref{xtp1}).}
\EndFor
\State {{\bf -- BS strategy (inner iteration):}}
\For { $\forall j\in\mathcal{S}_{0}$ }
\State {Calculate $K_{j}^{(t+1)}$'s according to (\ref{Kjp1}).}
\State {Update $\mu^{(t+1)}$ and $\nu^{(t+1)}$ according to (\ref{mutp1}) and (\ref{nutp1}).}
\EndFor
\EndWhile
\State {Updated cell association $\mathbf{X}^{\star(s+1)}$ as $\mathbf{X}^{(t+1)}$ obtained at convergence of inner iterations.}
\State {{\bf -- Small cell strategy (outer iteration):}}
\For { $\forall j\in\mathcal{S}$ }
\State {Calculate locally smallest feasible $\beta$ as}
\State {$\beta=\frac{\sum_{k\in\mathcal{U}}x_{j,k}^{\star(s+1)}r_{j,k}}{\sum_{k\in\mathcal{U}}x_{j,k}^{\star(s+1)}(r_{j,k}+c_{j})}$.}
\State {Feed back locally smallest feasible $\beta$ to the macro BS.}
\EndFor
\State {{\bf -- Macro BS strategy (outer iteration):}}
\State {Designate the maximum of the locally smallest feasible $\beta$s as $\beta^{(s+1)}$.}
\State {Pass $\beta^{(s+1)}$ to the inner iterations for the CA sub-problem.}
\EndWhile
\State{Return $\mathbf{X}^{\star(s+1)}$ and $\beta^{\star(s+1)}$ at convergence of outer iterations
as the optimal values $\mathbf{X}^{\star}$ and $\beta^{\star}$.}
\end{algorithmic}
\end{algorithm}

%%%%%%%%%%%%%%%%%%%%%%%%%%%%%%%%%%%%%%%%%%%%%%%%%%%%%%%%%%%%%%%%%%%%%%%%%%%%%%%%%%%%%%%%%%%%%%%%%%
\section{Joint CA-WBBA with Per-Small Cell Wireless Backhaul Bandwidth Allocation}\label{five}
According to (\ref{betau}), the unified WBBA factor $\beta$ studied in Section \ref{four} is chosen as
the maximum of the smallest feasible per-small-cell WBBA ratios.
Because different channel and load conditions are experienced by the small cells,
the minimum WBBA ratio for each individual small cell may vary a lot.
If the small cells within the macro cell range can adaptively use different WBBA factors,
better spectral efficiency can be achieved.
We are therefore motivated to investigate the p-WBBA scheme for joint CA-WBBA design in a
two-tier HetNet.

\subsection{Problem Formulation}\label{5a}

With p-WBBA, we no longer use a unified and dedicated frequency band for wireless backhauling
throughout the macro cell range.
Interference management in the HetNet therefore becomes more involved.
Similar to the problem formulation for the unified WBBA case in Section \ref{four},
we define the small cell throughput $R_{j}^{P}$, macro cell user rate $R_{0,k}^{P}$
and small cell DL backhaul capacity $C_{j}^{P}$ in the following,
where the superscript ``P''  indicates per-small-cell WBBA.

We dynamically allocate a fraction of the frequency band in a small cell $j$ for its wireless backhaul.
A per-small-cell frequency allocation factor $\beta_{j}\in[0,1]$ is introduced to denote
the fraction of bandwidth allocated for wireless backhaul in the $j^{\rm{th}}$ small cell.
The remaining bandwidth is used for the small cell's service functionality.
The throughput of the $j^{\rm{th}}$ small cell in the p-WBBA scenario is
\begin{equation}\label{RjC}
R_{j}^{P} = \frac{1-\beta_{j}}{\sum_{k\in\mathcal{U}}x_{j,k}}
\sum_{k\in\mathcal{U}}x_{j,k}\log_{2}(1+\gamma_{j,k}).
\end{equation}
%where $P_{j}$ is the average transmit SNR of the $j^{\rm{th}}$ small cell.
%Because of the wireless backhaul constraint, the throughput $R_{j}$ given by (\ref{Rj})
%must not exceed the achievable rate of the wireless communication link
%between the macro BS and the $j^{\rm{th}}$ small cell, which will be analyzed in the following.
%
%In the ``large-scale MIMO'' regime, the macro BS has an antenna array size $N_{T}$
%much greater than the number of simultaneously served nodes (including both MTs and the small cells).
%In this work, we assume $N_{T}\gg N_{g}$ and $N_{U}\gg N_{g}$.
%A large number of MTs associated with the macro BS is served by resource sharing in other dimensions,
%e.g., time sharing.
%When a mobile terminal $k$ is associated with the macro BS,
%its achievable rate $R_{0,k}$ in the ``large-scale MIMO'' regime is determined by both the channel $\bar{H}_{k}$
%and the number of MTs associated with the massive MIMO macro BS \cite{caire}.
Given the small cell bandwidth allocation factor $\beta_{j}$ for every $j\in\mathcal{S}$,
if the sub-frequency-band assignment in each small cell is done in a random fashion,
the average number of small cells associated with the macro BS on each sub-frequency-band for backhauling
is $N_{b}=\sum_{j\in\mathcal{S}}\beta_{j}$.
Consequently, similar to the unified WBBA case,
the achievable rate of the $k^{\rm{th}}$ MT in the macro BS under p-WBBA is given by
\begin{equation}\label{R0kC}
R_{0,k}^{P} = x_{0,k}\frac{N_{g}-N_{b}}{\sum_{k\in\mathcal{U}}x_{0,k}}
\log_{2}\left(1+\frac{N_{T}-N_{g}+1}{N_{g}}\gamma_{0,k}\right),
\end{equation}
where we still assume equal resource sharing among MTs associated with the macro BS.
The DL wireless backhaul capacity for small cell $j$ under p-WBBA is
\begin{equation}\label{CjC}
C_{j}^{P} = \beta_{j}\log_{2}\left(1+\frac{N_{T}-N_{g}+1}{N_{g}}\gamma_{j}\right).
\end{equation}

The system's sum of the logarithmic MT rates in case of p-WBBA is given by
\begin{equation}\label{logthroughputC}
%\begin{aligned}
\bar{R}^{P} = \sum_{k\in\mathcal{U}}x_{0,k}\log\left(\frac{1-\frac{\sum_{j\in\mathcal{S}}\beta_{j}}{N_{g}}}{\sum_{k\in\mathcal{U}}x_{0,k}}
%\log_{2}\left(1+\frac{N_{T}-N_{g}+1}{N_{g}}P_{0}\bar{H}_{k}\right)
\ r_{0,k}\right)+\sum_{j\in\mathcal{S}}\sum_{k\in\mathcal{U}}x_{j,k}
\log\frac{(1-\beta_{j})r_{j,k}}{\sum_{k\in\mathcal{U}}x_{j,k}}
%&= \sum_{k\in\mathcal{U}}x_{0,k}\bar{R}_{0,k} + \sum_{j\in\mathcal{S}}\sum_{k\in\mathcal{U}}x_{j,k}\bar{R}_{j,k}.
= \bar{R}_{0}^{P} + \sum_{j\in\mathcal{S}}\bar{R}_{j}^{P} = \sum_{j\in\mathcal{S}_{0}}\bar{R}_{j}^{P}.
%\end{aligned}
\end{equation}

Denote $\mathbf{b}=\{\beta_{j};\ j\in\mathcal{S}\}$ and the joint downlink CA-WBBA problem under per-small-cell WBBA
is formulated as the following problem $\mathbf{P2}$, which is also a MINLP problem.
\begin{eqnarray}%\label{opt2}
%\begin{array}{ll}
\mathbf{P2:}\ \ \underset{\mathbf{b},\mathbf{X}}{\mbox{maximize}}
                                &&\sum_{j\in\mathcal{S}_{0}}\bar{R}_{j}^{P}(\mathbf{b},\mathbf{X}) \nonumber\\
\label{constraint2.1}
\mbox{subject to}               &&x_{j,k}\in\{0,1\},\ \forall j\in\mathcal{S}_{0}, k\in\mathcal{U};\\
\label{constraint2.2}
\                               &&\sum_{j\in\mathcal{S}_{0}}x_{j,k}=1,\ \forall k\in\mathcal{U};\\
\label{constraint2.3}
\                               &&0\leq \beta_{j}\leq 1,\ \forall j\in\mathcal{S};\\
\label{constraint2.4}
\                               &&\beta_{j}\leq\sum_{k\in\mathcal{U}}x_{j,k},\ \forall j\in\mathcal{S};\\
\label{constraint2.5}
\                               &&R_{j}^{P}\leq C_{j}^{P},\ \forall j\in\mathcal{S}.
%\end{array}
\end{eqnarray}
Constraint (\ref{constraint2.4}) is used to assign $\beta_{j}=0$
when there is no MTs associated with small cell $j$, i.e., $\sum_{k\in\mathcal{U}}x_{j,k}=0$.

%In \textbf{Appendix \ref{Appendix-B}}, we show that

The objective function of $\mathbf{P2}$, which is specified in (\ref{logthroughputC}), is nonlinear and non-convex.
Moreover, (\ref{logthroughputC}) is not separable in the primal variables $\mathbf{b}$ and $\mathbf{X}$
as in the objective function of $\mathbf{P1}$.
Also, the wireless backhaul constraint in (\ref{constraint2.5}) is a nonlinear and non-convex coupling constraint.
As we can observe from the problem formulation $\mathbf{P2}$ and equations (\ref{RjC})-(\ref{CjC}),
with large-scale MIMO and p-WBBA for small cell backhauling,
the small cell specific bandwidth allocation factor $\beta_{j}$ not only determines backhaul and user rates
of small cell $j$, but also dynamically impacts the user rates in the macro BS.
A decomposition approach, which decouples the design variables with unified WBBA in Section \ref{4b}, is not directly applicable.
The  constraint in (\ref{constraint2.5}) is a nonlinear coupling constraint,
which can be made linear if we conduct cell association and wireless backhaul bandwidth allocation separately.

In this section, in addition to a distributed algorithm similar to that proposed in Section \ref{four},
we investigate centralized fast heuristic algorithms for joint downlink CA-WBBA problem
under p-WBBA.
Directly solving the MINLP problem by the BONMIN (Basic Open-source Nonlinear Mixed INteger programming) solver \cite{bonmin}
on the GAMS platform \cite{gams} is also considered
to provide benchmarks for the numerical results in Section \ref{six}.

\subsection{Separate Cell Association and Per-Small Cell Bandwidth Allocation for Small Cell Backhauling}\label{5b}

By examining problem $\mathbf{P2}$, we observe that with per-small-cell WBBA,
the wireless backhaul constraint (\ref{constraint2.5}) is an active constraint
so the optimum must be achieved at equality of (\ref{constraint2.5}) for all the small cells.
Given the cell association indicators $\{x_{j,k};(j,k)\in\mathcal{S}_{0}\times\mathcal{U}\}$,
the optimal bandwidth allocation factor $\beta_{j}$ can thus be determined locally at each small cell
by equating (\ref{RjC}) with (\ref{CjC}).
With distributed cell association, $x_{j,k}$s can be determined at each MT, as in Section \ref{4c}.
Centralized cell association, on the other hand, relies on a central scheduler,
which can be located at the macro BS or at the radio network controller (RNC).

A straightforward solution to the joint downlink CA-WBBA problem is to
consider the two tasks separately and iteratively in two algorithm stages,
which is similar to the hierarchical decomposition approach investigated in Section \ref{four}.
Given the cell association variables, which can be obtained by any simple CA strategy,
e.g., SINR-based cell association,
the optimal bandwidth allocation factors $\beta_{j}$ conditioned on the cell association
can be determined accordingly with the knowledge of the wireless channel states as
\begin{equation}\label{betaj}
\beta_{j}=\frac{\sum_{k\in\mathcal{U}}x_{j,k}r_{j,k}}{\sum_{k\in\mathcal{U}}x_{j,k}(r_{j,k}+c_{j})}.
\end{equation}

The small cell WBBA factors $\beta_{j}$s obtained from (\ref{betaj}) are fed back to
the cell association sub-problem to update the cell association scheme.
%According to (\ref{B.1}),
The utility function $\bar{R}^{P}$ to be maximized is concave given that $\beta_{j}$s are constant.
Therefore, similar to the dual decomposition solution for sub-problem $\mathbf{P1.1r}$,
a distributed cell association algorithm based on Lagrangian dual decomposition can be employed here
to solve the cell association sub-problem distributively in each iteration of the master problem $\mathbf{P2}$.
Specifically, a Lagrangian similar to (\ref{lagrU}) is formulated for the CA sub-problem with per-small-cell WBBA
\begin{equation}\label{lagrC}
%\begin{aligned}
\mathcal{L}^{P}(\mathbf{X},\mu,\nu)
= \sum_{k\in\mathcal{U}}\sum_{j\in\mathcal{S}_{0}}x_{j,k}\left[d_{j}+\log(r_{j,k})-\mu_{j}-\nu_{j}(1-\beta_{j})r_{j,k}\right]
 + \sum_{j\in\mathcal{S}_{0}}K_{j}\left[\mu_{j}-\log(K_{j})+\nu_{j}\beta_{j}c_{j}\right],
%\end{aligned}
\end{equation}
where we have defined the per-small-cell WBBA related terms
$d_{0}=\log\left(1-\frac{\sum_{j\in\mathcal{S}}\beta_{j}}{N_{g}}\right)$
and $d_{j}=\log(1-\beta_{j})$ for $j\in\mathcal{S}$.
The subsequent dual problem formulation is the same as $\mathbf{D1.1}$,
except that $g_{k}(\mu,\nu)$ and $g_{K}(\mu,\nu)$ are slightly modified to include $d_{j}$s
and use per-small-cell WBBA factors $\beta_{j}$s instead of a unified $\beta$.
Specifically, we substitute $(\log(r_{j,k})-\mu_{j}-\nu_{j}(1-\beta)r_{j,k})$ in (\ref{gkmu})
by $(d_{j}+\log(r_{j,k})-\mu_{j}-\nu_{j}(1-\beta_{j})r_{j,k})$,
and simply replace $\beta$ in (\ref{gKKmu}) by $\beta_{j}$.
A distributed algorithm based on Lagrangian dual decomposition similar to that
in Section \ref{4c} can be used in the p-WBBA scenario
for the cell association sub-problem.

\subsection{Macro BS Offloading and Small Cell Load Balancing}\label{5c}

The use of per-small-cell WBBA factors inevitably increases the amount of message passing
for the distributed algorithm discussed in Section \ref{5b}.
The corresponding system overhead may be an issue in practical communication scenarios.
Especially separating the CA and WBBA stages for problem $\mathbf{P2}$ is not based on
decoupling of CA and WBBA variables as in the upper-level primal decomposition for $\mathbf{P1}$.
The distributed iterative algorithm in Section \ref{5b} is sub-optimal.
In this subsection, we consider designing fast heuristic algorithm based on the idea of BS offloading
which helps to achieve satisfactory performance.
Such a scheme should provide a good compromise between performance and overhead/complexity.

We start with %the simplest cell association strategy, i.e.,
the SINR-based cell association.
Other cell association techniques can also be employed at the initial cell association stage.
In SINR-based cell association, each MT selects the cell which achieves the strongest SINR
as determined in Section \ref{three}.
Then in each small cell, the wireless backhaul bandwidth allocation factor is calculated
locally according to (\ref{betaj}).
With SINR-based cell association, the high power macro BS attracts more MTs and
subsequently causes load imbalance between the two tiers.
Some small cells may not make good use of their available wireless backhaul resources
because of their small or even zero load,
while some are overloaded due to heavy load or less favorable channel to the macro BS.
We are therefore motivated to design a heuristic algorithm for joint CA-WBBA from the perspective of
load balancing in a two-tier HetNet
where large-scale MIMO at the macro BS and wireless backhauling for the small cells come into play.
%based on the cell association and wireless backhaul bandwidth allocation algorithm in Section \ref{4b}.

We first consider offloading MTs associated with the macro BS to small cells for sum log-rate maximization.
A greedy algorithm is used to sequentially check each MT associated with the macro BS
to see if an improved utility can be achieved by offloading it to a small cell.
The details of the offloading algorithm are given in \textbf{Algorithm \ref{alg1}}.
It can be shown that the complexity of this algorithm is of $\mathcal{O}(N_{U}N_{S})$.
%Alg. \ref{alg1} is iterated until convergence.
\begin{algorithm}
\caption{Offloading of the macro BS}\label{alg1}
\begin{algorithmic}
\State {{\bf Input:}  SINR-based cell association $\mathbf{X}$ and the subsequent wireless backhaul
resource allocation $\mathbf{b}$.
Large-scale channel SINR $\{\gamma_{j,k}; j\in{\mathcal{S}}_{0}, k\in\mathcal{U}\}$ and $\{\gamma_{j}; j\in\mathcal{S}\}.$}
\State {{\bf Initialization:} Calculate $N_{b}$, $\bar{R}_{0}^{P}$, $\bar{R}_{j}^{P}, \forall j\in\mathcal{S}$ and $\bar{R}^{P}$.}
\State {{\bf Output:} Offloaded cell association $\mathbf{X}^{\star}$.}
\State {{\bf Execution:}}
\For { $\forall k\in\mathcal{U}$ }
\If {( $x_{0,k}=1$ )}
\For {$j\in\mathcal{S}$}
\State {Let $\mathbf{X}_{T}=\mathbf{X}$ and set $x_{0,k}=0$, $x_{j,k}=1$ in $\mathbf{X}_{T}$.}
\State {Calculate new $\beta_{j}$ as $\beta'_{j}$ using $\mathbf{X}_{T}$ and (\ref{betaj}), and $N'_{b}=N_{b}-\beta_{j}+\beta'_{j}$.}
\State {Calculate new $\bar{R}^{P'}$ by updating $\bar{R}^{P'}_{0}$ and $\bar{R}^{P'}_{j}$ with $\beta'_{j}$, $N'_{b}$, $\gamma_{0,k}$, $\gamma_{j,k}$ and $\gamma_{j}$.}
\If {Improved $\bar{R}^{P'}>\bar{R}^{P}$ is observed,}
\State {Update $\beta_{j}=\beta'_{j}$, $N_{b}=N'_{b}$, $\bar{R}^{P}_{0}=\bar{R}^{P'}_{0}$, $\bar{R}^{P}_{j}=\bar{R}^{P'}_{j}$, $\bar{R}^{P}=\bar{R}^{P'}$ and $\mathbf{X}=\mathbf{X}_{T}$.}
\EndIf
\EndFor
\EndIf
\EndFor
\State{Return the updated $\mathbf{X}$ as $\mathbf{X}^{\star}$.}
\end{algorithmic}
\end{algorithm}

By running the same offloading procedure for each non-empty small cell after the macro BS offloading,
improvement in  the utility $\bar{R}^{P}$ is expected.
An overall complexity of $\mathcal{O}(N_{U}N_{S}^{2})$ will be introduced by this supplementary greedy algorithm.
However, the numerical studies %(as will be presented in the next section)
revealed that the improvement due to
the extra small cell load balancing procedure is negligible.

%%%%%%%%%%%%%%%%%%%%%%%%%%%%%%%%%%%%%%%%%%%%%%%%%%%%%%%%%%%%%%%%%%%%%%%%%%%%%%%%%%%%%%%%%%%%%%%%%%
\section{Numerical Results and Discussion}\label{six}

\begin{figure}[!htb]
\minipage{0.49\textwidth}
  \includegraphics[width=3.2in]{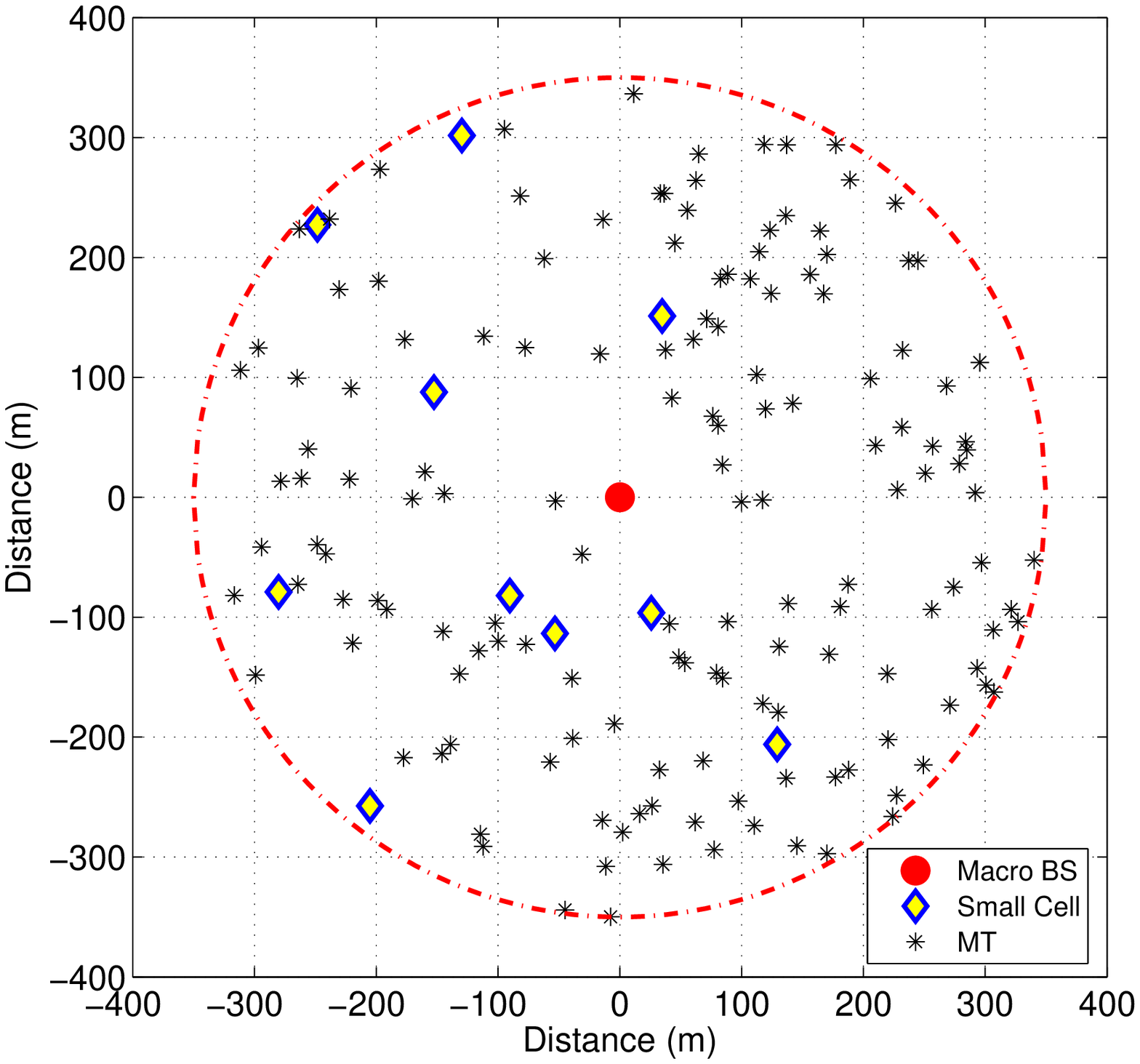}
 \caption{An example of the network geometry used for the simulation.}
\label{nettopo}
\endminipage\hfill
\minipage{0.49\textwidth}
  \includegraphics[width=3.2in]{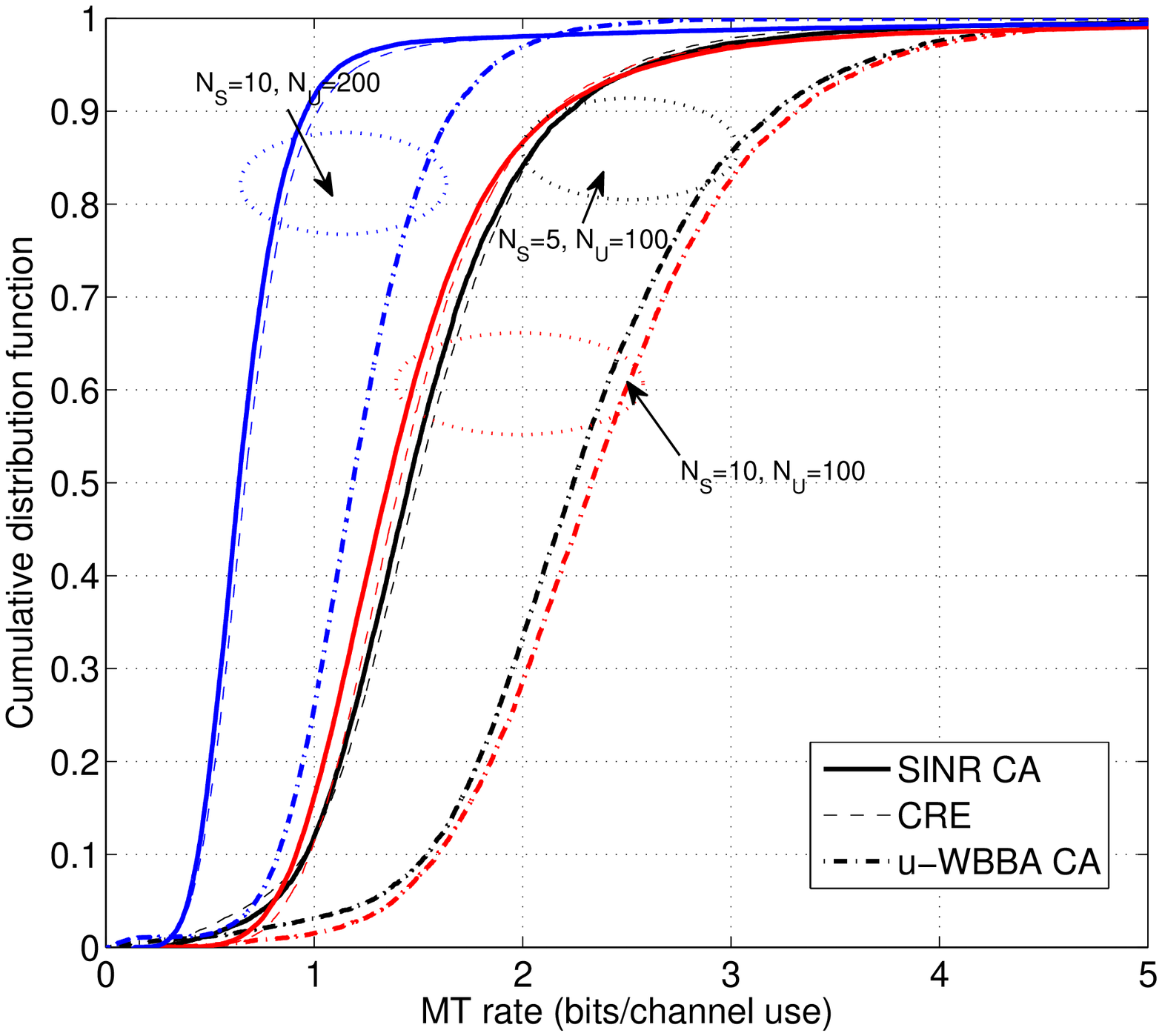}
\caption{Empirical CDFs of the MT rate  for different cell association schemes under the u-WBBA scenario.}
\label{eCDF_u}
\endminipage\hfill
\end{figure}

In this section, we examine the proposed joint downlink cell association and wireless backhaul
bandwidth allocation schemes for the two-tier large-scale MIMO HetNet through numerical examples.
In each simulation trial, we simulate one large-scale MIMO macro BS at the center of its cell range,
which is a circle of radius 350 m.
$N_{S}$ small cells and $N_{U}$ mobile terminals are randomly and uniformly placed in the circular macro cell range.
An example (random realization) of the network geometry for the simulation with $N_{S}=10$ small cells
and $N_{U}=100$ mobile terminals is shown in Fig. \ref{nettopo}.

%%%%%%%%%%%%%%%%%%%%%%%%%%%%%%%%%%%%%%%%%%%%%%%%%%%%%%%%%%%%%%%%%%%%%%%%%%%%%%%%%%%%%%%%%%%%%%%%%%%%%%%%%%%%%%%%
%\begin{figure}
%\begin{center}
%\includegraphics[width=4.5in, draft=false]{nettopo.eps}
%\caption{An example of the network geometry used for the simulation.}
%\label{nettopo}
%\end{center}
%%\vskip -10pt
%\end{figure}
%%%%%%%%%%%%%%%%%%%%%%%%%%%%%%%%%%%%%%%%%%%%%%%%%%%%%%%%%%%%%%%%%%%%%%%%%%%%%%%%%%%%%%%%%%%%%%%%%%%%%%%%%%%%%%%%

The macro BS has a transmit power of 43 dBm (20 W) and the small cells have transmit power of 33 dBm (2W).
%The thermal noise power for a 5 MHz bandwidth is $N_{0}=-107$ dBm.
In addition, we assume a 5 dB transmit and receive antenna gain for the small cells,
as opposed to 0 dB gain of the mobile terminals.
The large-scale path-loss and shadowing model as described in Section \ref{2b} is employed.
At the massive MIMO macro BS, an $N_{T}=100$ antenna array with beamforming group size of $N_{g}=20$ is considered.
Different $N_{U}$ and $N_{S}$ values are examined in the numerical examples for the proposed algorithms.
The SINR-based cell association and the results for CRE of small cells \cite{CRE}
are also evaluated under the small cell wireless backhaul constraint for comparison.
The SINR bias used in the numerical examples for the CRE scheme is 3 dB.
For each set of system settings, the numerical results are obtained by averaging over
a sufficiently large number of simulation trials.

\subsection{Joint CA-WBBA With Unified WBBA Factor}\label{6a}
The empirical cumulative distribution function (CDF) of the long-term MT rate in the two-tier HetNet
employing the proposed distributed joint CA-WBBA algorithm for the u-WBBA scenario is shown in Fig. \ref{eCDF_u}.
The CDFs of downlink transmission rates for MTs with the SINR-based cell association and CRE schemes are also plotted.
In order to study the impact of the small cell density and the density of MTs on distribution of the transmission rates for the MTs,
the CDFs for different $N_{S}$ and $N_{U}$ values are considered.

%%%%%%%%%%%%%%%%%%%%%%%%%%%%%%%%%%%%%%%%%%%%%%%%%%%%%%%%%%%%%%%%%%%%%%%%%%%%%%%%%%%%%%%%%%%%%%%%%%%%%%%%%%%%%%%%
%\begin{figure}
%\begin{center}
%\includegraphics[width=4.5in, draft=false]{ecdf_uWBBA.eps}
%\caption{Empirical CDFs of the MT rate in the two-tier HetNet for different cell association schemes under the
%unified WBBA constraint of small cells.}
%\label{eCDF_u}
%\end{center}
%\vskip -10pt
%\end{figure}
%%%%%%%%%%%%%%%%%%%%%%%%%%%%%%%%%%%%%%%%%%%%%%%%%%%%%%%%%%%%%%%%%%%%%%%%%%%%%%%%%%%%%%%%%%%%%%%%%%%%%%%%%%%%%%%%

\begin{figure}
\centering
\subfigure[]{
   \includegraphics[width=3.45in, draft=false] {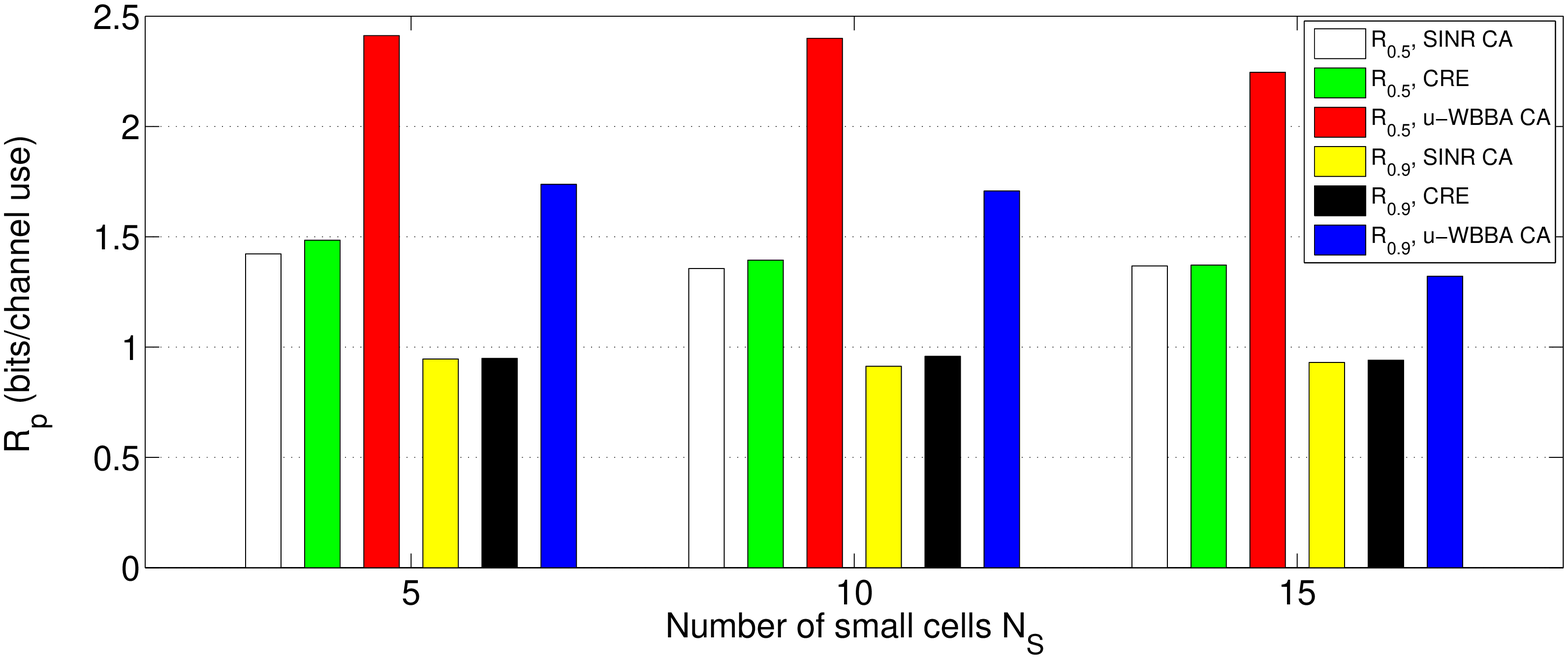}
   \label{MTrate_u1}
 }
 \subfigure[]{
   \includegraphics[width=3.45in, draft=false] {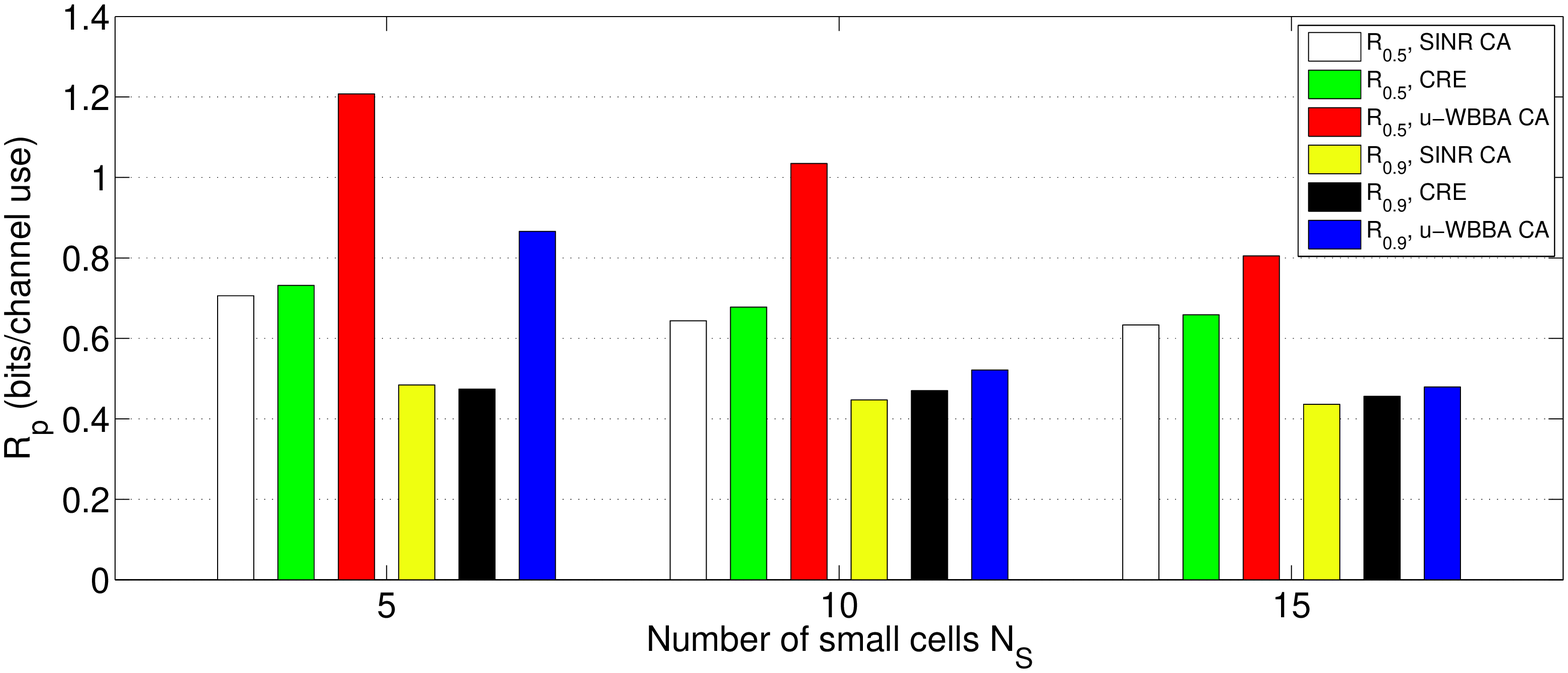}
   \label{MTrate_u2}
 }
\caption{The $R_{0.5}$ and $R_{0.9}$ MT rate values for various cell association schemes with different
$N_{S}$ values and (a) $N_{U}=100$ MTs and (b) $N_{U}=200$ MTs under u-WBBA.}
\label{MTrate_u}
\vskip-8pt
\end{figure}

It is observed from Fig. \ref{eCDF_u} that under the u-WBBA scenario, the proposed
distributed joint CA-WBBA algorithm has the CDF curves significantly shifted towards the right,
compared with the SINR-based cell association,
which is an indication of improved MT rate for most of the MTs in the network.
To better demonstrate the improvement in overall MT rate achieved by the proposed joint CA-WBBA scheme,
we define the MT rate at probability $p$, $R_{p}$, in (\ref{Rp}) for $p\in[0,1]$ and show in Fig. \ref{MTrate_u}
the values of $R_{0.5}$ and $R_{0.9}$ for different cell association strategies in different system settings.
\begin{equation}\label{Rp}
\Pr\{\mbox{MT rate}>R_{p}\}=p.
\end{equation}
$R_{0.5}$ is the median MT rate, and $R_{0.9}$ is the rate that can be attained by 90\% of the MTs in the network.
Both of these parameters can be considered as indicators of the overall QoS provisioning of the two-tier HetNet.
%%%%%%%%%%%%%%%%%%%%%%%%%%%%%%%%%%%%%%%%%%%%%%%%%%%%%%%%%%%%%%%%%%%%%%%%%%%%%%%%%%%%%%%%%%%%%%%%%%%%%%%%%%%%%%%%
%%%%%%%%%%%%%%%%%%%%%%%%%%%%%%%%%%%%%%%%%%%%%%%%%%%%%%%%%%%%%%%%%%%%%%%%%%%%%%%%%%%%%%%%%%%%%%%%%%%%%%%%%%%%%%%%

According to the QoS indicators $R_{0.5}$ and $R_{0.9}$, the distributed joint CA-WBBA algorithm
significantly outperforms the other cell association schemes under u-WBBA.
Consider the SINR-based cell association as the baseline scheme.
Up to 77\% improvement in $R_{0.5}$ and 87\% improvement in $R_{0.9}$ are achieved by the proposed algorithm
with $N_{S}=10$ and $N_{U}=100$.
The improvement in the QoS indicators by the proposed joint CA-WBBA algorithm is consistently around 60\% in most of the scenarios examined.
However, the CRE of small cells by SINR biasing provides negligible improvement over SINR-based cell association,
which is obvious in both the CDF curves (in Fig. \ref{eCDF_u}) and the corresponding $R_{p}$ values (in Fig. \ref{MTrate_u}).
Therefore, the CRE is not an efficient solution for cell association in presence of in-band wireless backhauling for the small cells.

%%%%%%%%%%%%%%%%%%%%%%%%%%%%%%%%%%%%%%%%%%%%%%%%%%%%%%%%%%%%%%%%%%%%%%%%%%%%%%%%%%%%%%%%%%%%%%%%%%%%%%%%%%%%%%%%
%\begin{figure}
%\begin{center}
%\includegraphics[width=4.5in, draft=false]{lograte_u.eps}
%\caption{Sum of the logarithmic MT rates of the proposed distributed cell association algorithm
%under u-WBBA constraint with different small cell and MT densities.
%The performance of the SINR-based and CRE cell association schemes are evaluated for comparison.}
%\label{lograte_u}
%\end{center}
%\vskip -10pt
%\end{figure}
%%%%%%%%%%%%%%%%%%%%%%%%%%%%%%%%%%%%%%%%%%%%%%%%%%%%%%%%%%%%%%%%%%%%%%%%%%%%%%%%%%%%%%%%%%%%%%%%%%%%%%%%%%%%%%%%

\begin{figure}[!htb]
\minipage{0.49\textwidth}
  \includegraphics[width=3.2in]{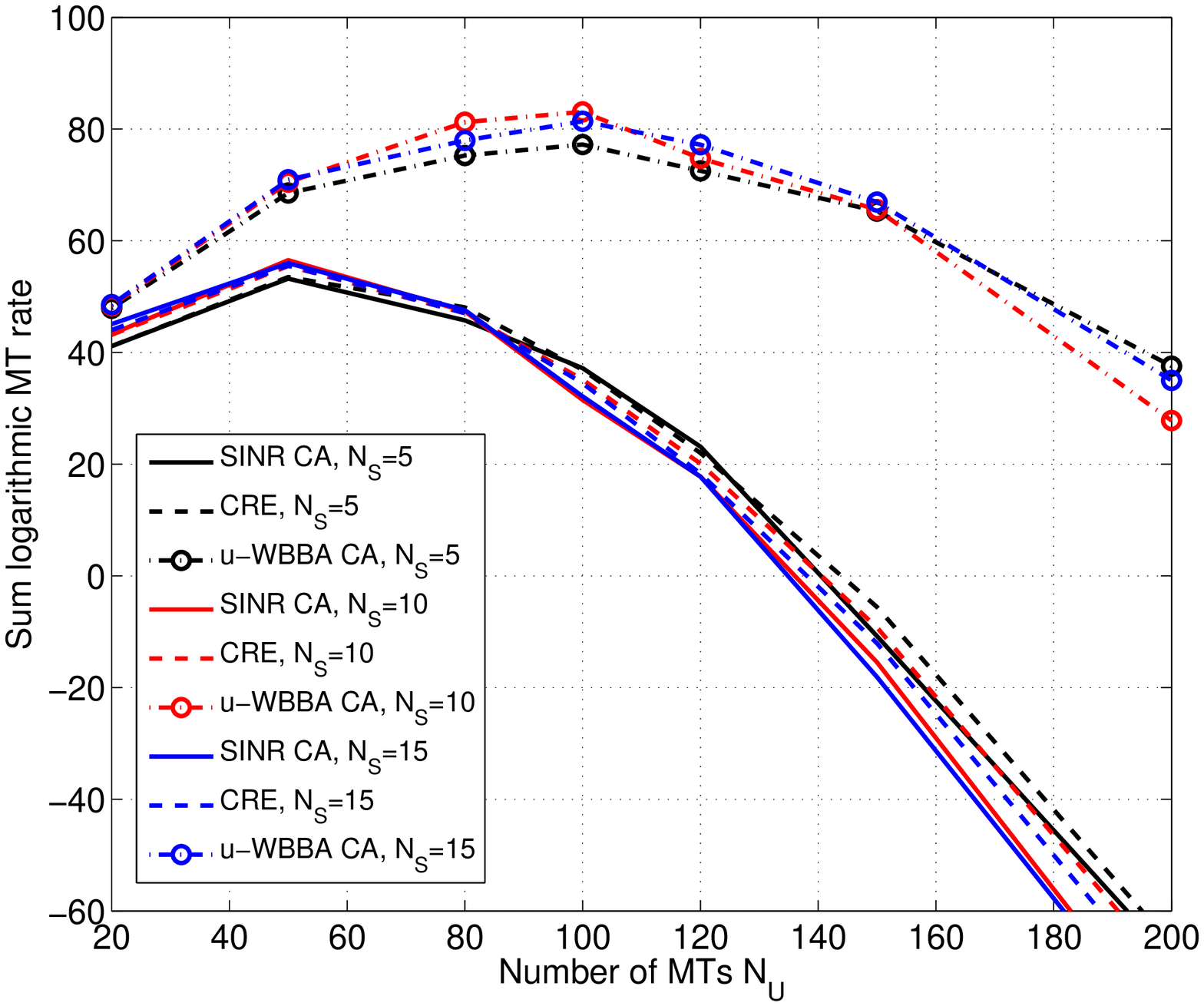}
 \caption{Sum of the logarithmic of MT rates for the proposed distributed cell association algorithm
under u-WBBA with different densities of small cells and MTs.
The performance of the SINR-based and CRE cell association schemes are shown for comparison.}
\label{lograte_u}
\endminipage\hfill
\minipage{0.49\textwidth}
 \includegraphics[width=3.6in]{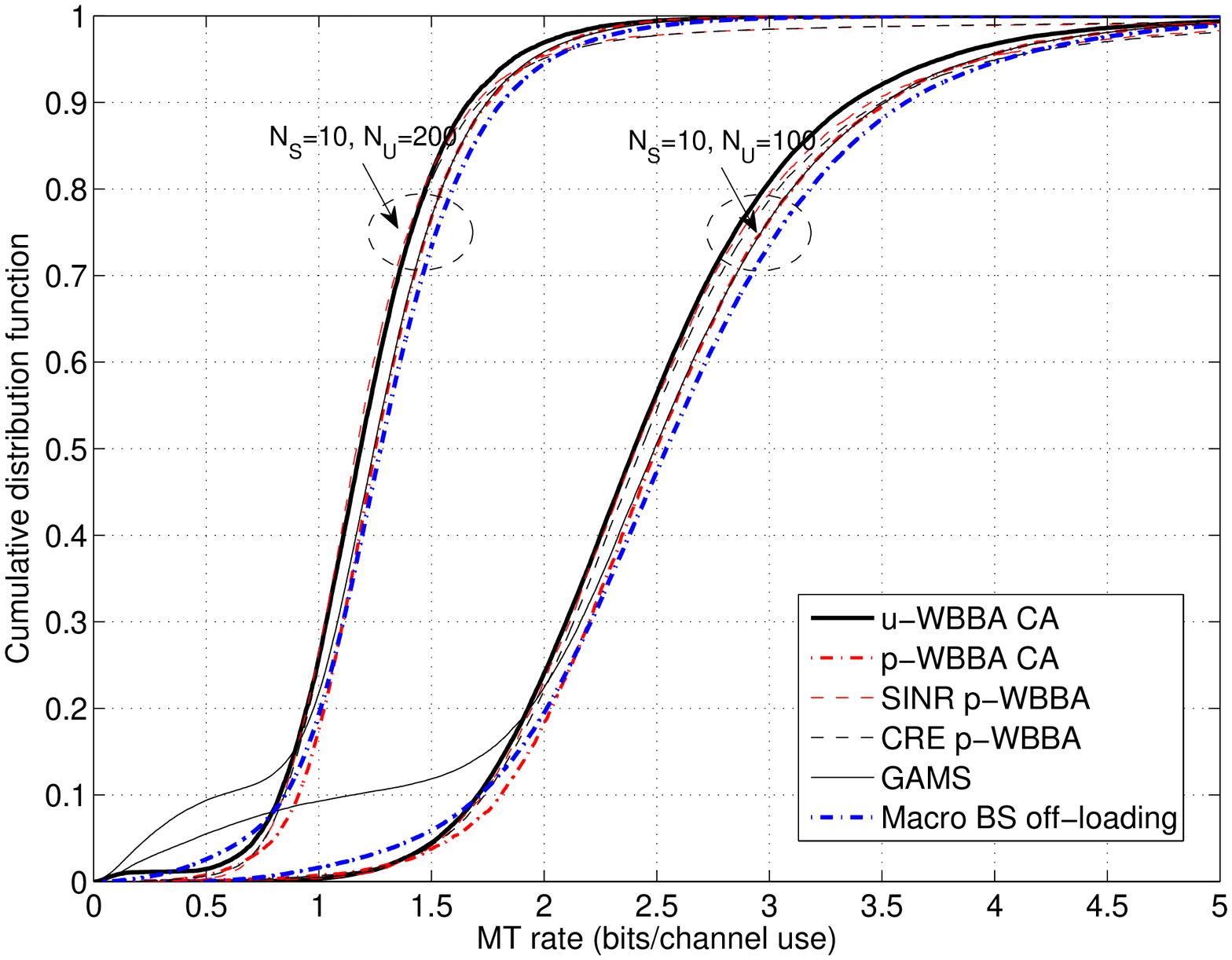}
\caption{Empirical CDFs of the MT rate in the two-tier HetNet for different cell association schemes under the p-WBBA constraint of small cells.}
\label{eCDF_p}
\endminipage\hfill
\end{figure}

We then evaluate through simulations the performance of different cell association strategies in terms of
the objective function value of $\mathbf{P1}$.
The sum of the logarithmic rates of all the MTs in the network is shown in Fig. \ref{lograte_u} for different system settings.
The superiority of the joint CA-WBBA algorithm based on hierarchical decomposition is still phenomenal,
%Again, significant improvement in the objective function value is achieved by the proposed joint CA-WBBA algorithm under u-WBBA constraint,
while applying the CRE strategy gives very small gain over the SINR-based cell association.
The objective function value does not change much with $N_{S}$, i.e., the small cell density.
This is a result of the dedicated in-band wireless backhaul design of the small cells,
as all the small cells have to share the same resource pool for backhauling.
Another interesting observation in Fig. \ref{lograte_u} is that an optimal system load (i.e., number of MTs) in terms of the objective function value
is located at about $N_{U}=100$ MTs for the proposed joint CA-WBBA algorithm, regardless of the small cell density.
On the other hand, the peak appears at around $N_{U}=50$ for the SINR-based cell association and the CRE strategy.
This implies that the proposed joint CA-WBBA algorithm is capable of taking full advantage of the multi-user diversity gain
of the macro BS antenna array with $N_{T}=100$ antenna elements, while the SINR-based and the CRE cell association schemes
can only enjoy part of benefits of the spatial degrees of freedom offered by the macro BS antenna array.

%%%%%%%%%%%%%%%%%%%%%%%%%%%%%%%%%%%%%%%%%%%%%%%%%%%%%%%%%%%%%%%%%%%%%%%%%%%%%%%%%%%%%%%%%%%%%%%%%%%%%%%%%%%%%%%%
%\begin{figure*}
%\begin{center}
%\includegraphics[width=6.5in, draft=false]{ecdf_pWBBAw.eps}
%\caption{Empirical CDFs of the MT rate in the two-tier HetNet for different cell association schemes under the
%p-WBBA constraint of small cells.}
%\label{eCDF_p}
%\end{center}
%%\vskip -10pt
%\end{figure*}
%%%%%%%%%%%%%%%%%%%%%%%%%%%%%%%%%%%%%%%%%%%%%%%%%%%%%%%%%%%%%%%%%%%%%%%%%%%%%%%%%%%%%%%%%%%%%%%%%%%%%%%%%%%%%%%%

\subsection{Joint CA-WBBA With Per-Small Cell WBBA Factors}\label{6b}
In this subsection we show numerical examples for the p-WBBA scenario.
Performances of the proposed algorithms, including the distributed joint CA-WBBA algorithm based on decomposition methods
and the centralized heuristic algorithm for macro BS offloading (OFL), are studied.
The SINR-based cell association and the CRE strategy are evaluated for comparison.
The solution given by the GAMS BONMIN solver is provided as a benchmark.
Performance of the small cell load balancing algorithm, which gives negligible improvement over the macro BS offloading algorithm,
is not shown in the results.

\begin{figure}
\centering
\subfigure[]{
   \includegraphics[width=3.45in, draft=false] {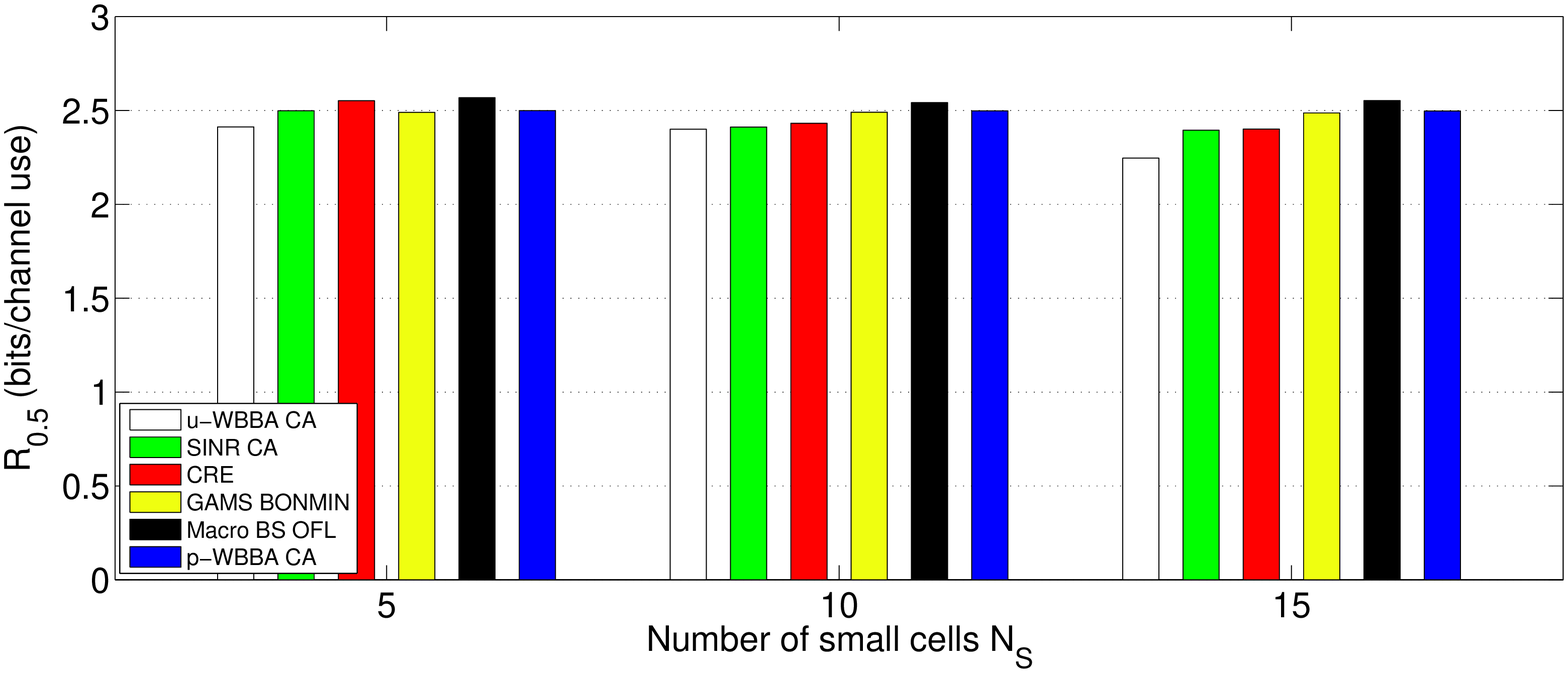}
   \label{MTrate_p5}
 }
 \subfigure[]{
   \includegraphics[width=3.45in, draft=false] {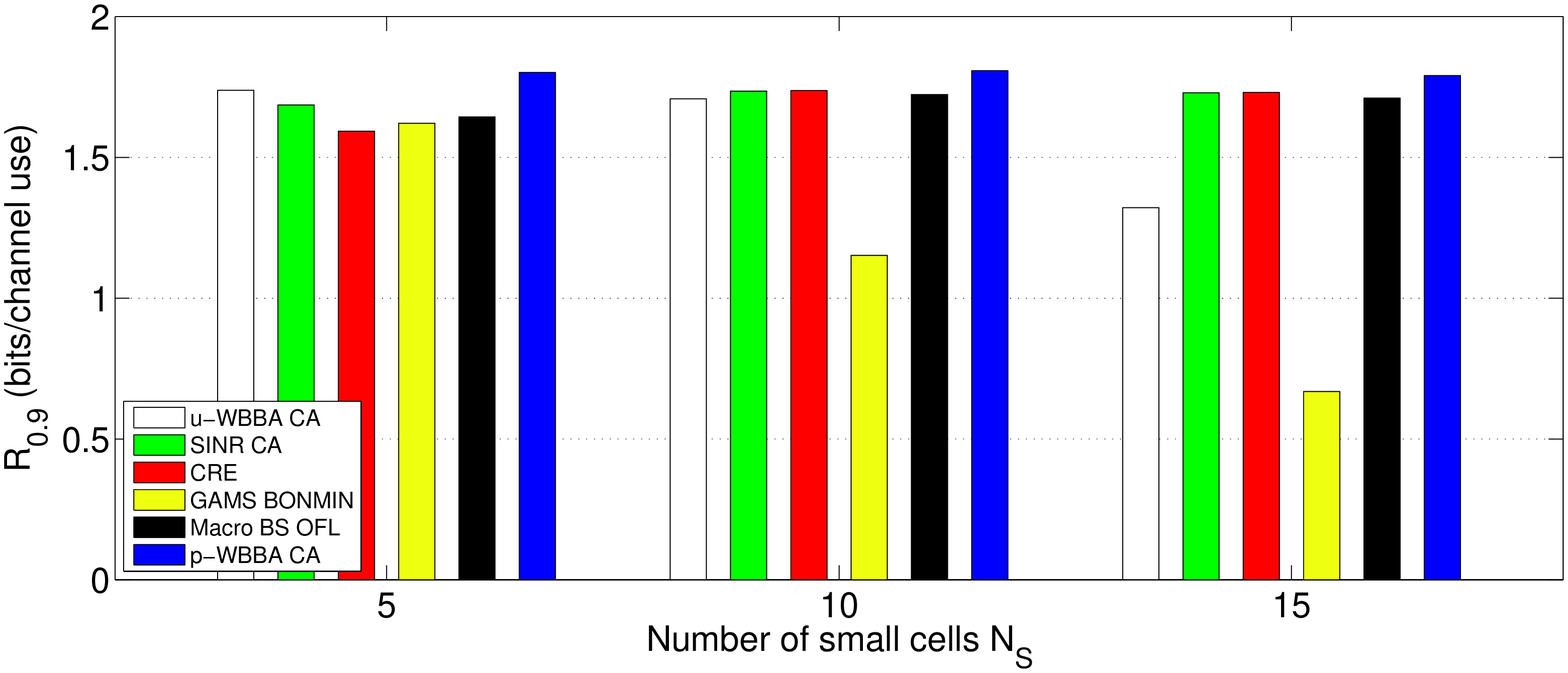}
   \label{MTrate_p9}
 }
\caption{The $R_{0.5}$ and $R_{0.9}$ MT rate values of various cell association schemes with
$N_{S}=10$ and $N_{U}=100$, under p-WBBA.}
\label{MTrate_p}
\vskip-8pt
\end{figure}

We show in Fig. \ref{eCDF_p} the empirical CDFs of the MT rate for different cell association schemes.
In order to focus on the main findings and make the plot clear, we only show the results for
$N_{S}=10$ with two different $N_{U}$ values ($N_{U}=100$ and $N_{U}=200$).
By exploiting the flexibility in spectral utilization of the per-small-cell WBBA strategy,
the rate performance of the MTs is in general improved compared with the unified WBBA scenario.
Among all the cell association schemes considered in the simulation for the p-WBBA scenario,
the distributed joint CA-WBBA algorithm and the centralized heuristic algorithm for macro BS offloading achieved
more desirable performance.
Specifically, the joint CA-WBBA algorithm can better reduce the fraction of MTs in the very low rate region,
while the macro BS offloading algorithm works best in increasing the probability that the MTs fall in the high rate region.
This can be interpreted as a tradeoff between fairness and median MT rate.
The CRE strategy again fails to achieve a significant gain over the SINR-based cell association.
The GAMS BONMIN solver, although gives very similar CDF curves as the joint CA-WBBA algorithm in the medium to high rate region,
sacrifices a relatively large fraction of the least favorable MTs by leaving them in the very low rate region.
Therefore, it is less preferable in terms of fairness.
Similar observations can be made from the comparison of the QoS indicators $R_{0.5}$ and $R_{0.9}$ in Fig. \ref{MTrate_p}
for $N_{S}=10$ and $N_{U}=100$.
However, in the p-WBBA scenario, the difference between CA schemes in $R_{0.5}$ and $R_{0.9}$ values is less significant
compared with the results for the u-WBBA scenario in Fig. \ref{MTrate_u}.

%%%%%%%%%%%%%%%%%%%%%%%%%%%%%%%%%%%%%%%%%%%%%%%%%%%%%%%%%%%%%%%%%%%%%%%%%%%%%%%%%%%%%%%%%%%%%%%%%%%%%%%%%%%%%%%%
%%%%%%%%%%%%%%%%%%%%%%%%%%%%%%%%%%%%%%%%%%%%%%%%%%%%%%%%%%%%%%%%%%%%%%%%%%%%%%%%%%%%%%%%%%%%%%%%%%%%%%%%%%%%%%%%

%%%%%%%%%%%%%%%%%%%%%%%%%%%%%%%%%%%%%%%%%%%%%%%%%%%%%%%%%%%%%%%%%%%%%%%%%%%%%%%%%%%%%%%%%%%%%%%%%%%%%%%%%%%%%%%%
\begin{figure}
\centering
\subfigure[]{
   \includegraphics[width=2in, height = 2.6in, draft=false] {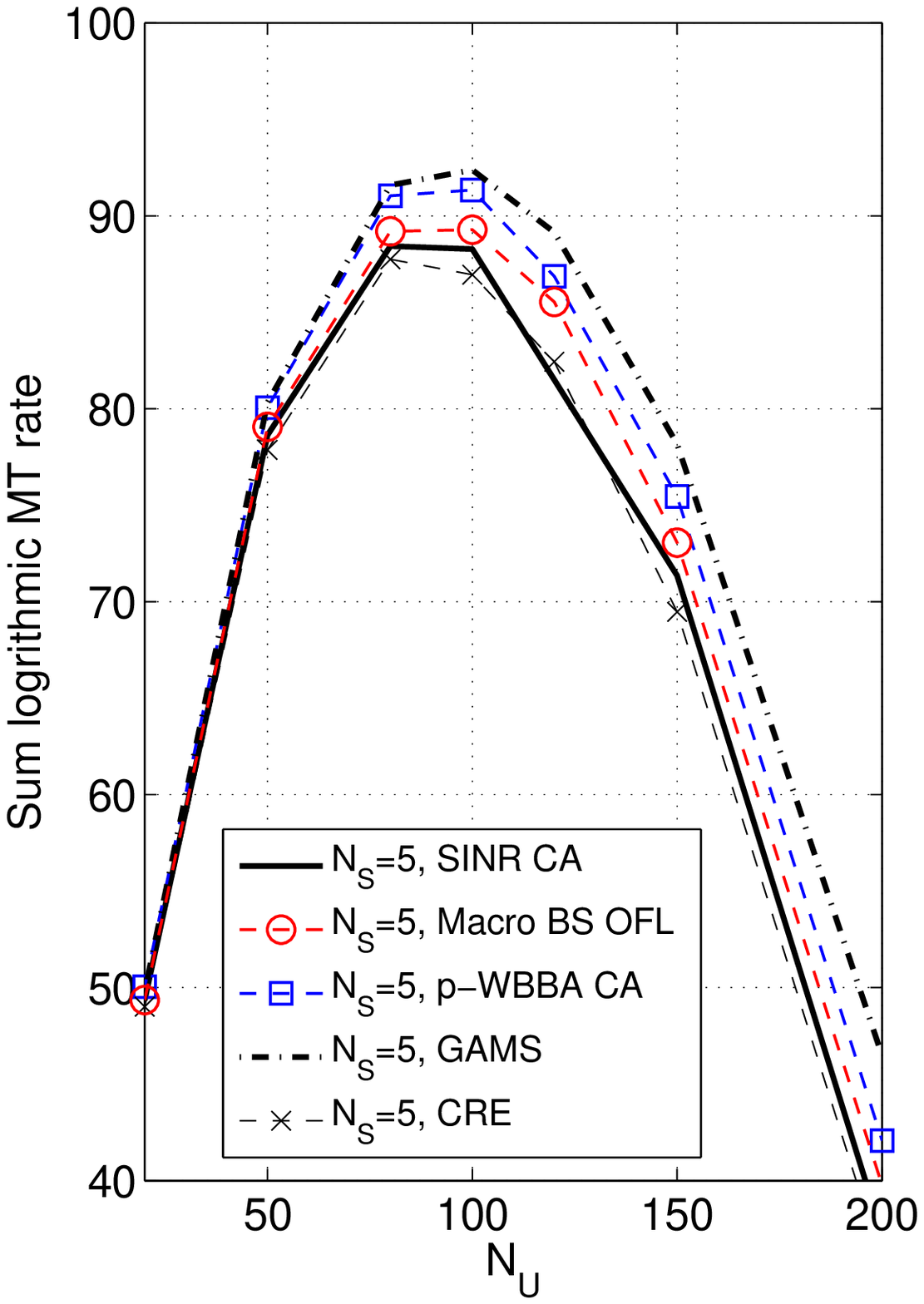}
   \label{lograte_p5}
 }
 \subfigure[]{
   \includegraphics[width=2in, height = 2.6in, draft=false] {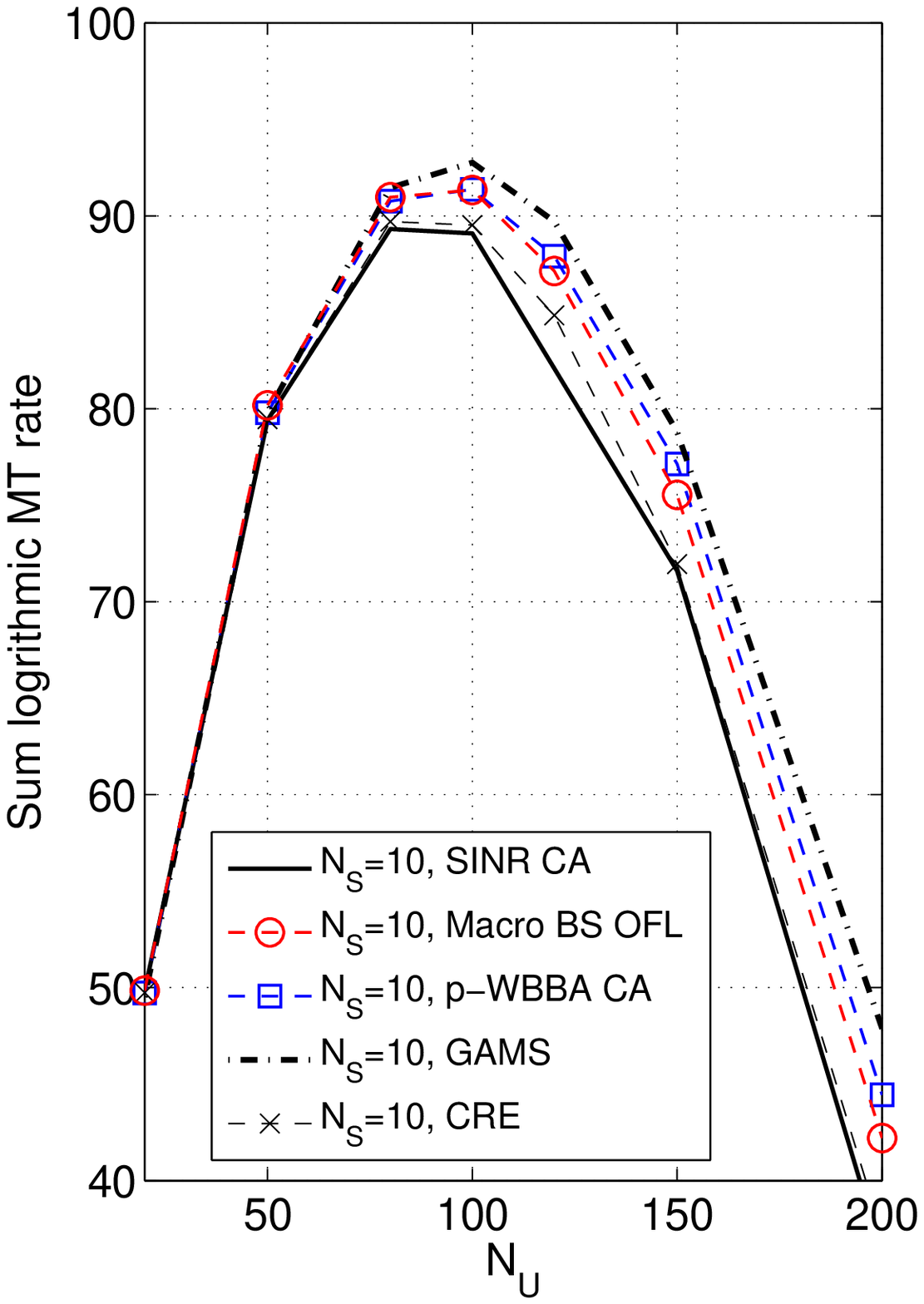}
   \label{lograte_p10}
 }
\caption{The $R_{0.5}$ and $R_{0.9}$ MT rate values of various cell association schemes with
$N_{S}=10$ and $N_{U}=100$, under p-WBBA.}
\label{lograte_p}
\vskip-8pt
\end{figure}
%%%%%%%%%%%%%%%%%%%%%%%%%%%%%%%%%%%%%%%%%%%%%%%%%%%%%%%%%%%%%%%%%%%%%%%%%%%%%%%%%%%%%%%%%%%%%%%%%%%%%%%%%%%%%%%%

The sum log-rate achieved by the distributed joint CA-WBBA algorithm and the heuristic algorithm
for macro BS offloading with different MT densities is shown in Figs. \ref{lograte_p5} and \ref{lograte_p10}
for $N_{S}$ = 5 and $N_{S}$ = 10, respectively, under p-WBBA.
The average utility values achieved by the SINR-based cell association, the CRE strategy, and the GAMS BONMIN solution are shown for comparison.
The GAMS BONMIN solution achieves the highest objective function value, which is the ``best'' from the optimization perspective.
However, because proportional fairness is considered in the optimization objective,
the solution still has fairness issues, which coincides with observations from Figs. \ref{eCDF_p} and \ref{MTrate_p}.
The distributed joint CA-WBBA algorithm based on the decomposition method outperforms all the other schemes
except the GAMS BONMIN solution in terms of the achieved objective function values.
By comparing Fig. \ref{lograte_p} with Fig. \ref{lograte_u}, it is worth noticing that employing the per-small-cell WBBA strategy
improves the objective function value.
Similar to the observations from Fig. \ref{lograte_u}, here we also see peaks of the curves locating at around $N_{U}=100$.
The difference is that, under p-WBBA, the SINR-based cell association and the CRE strategy
can also take full advantage of the multi-user diversity gain by having an optimal load approximately at $N_{U}=N_{T}=100$.

The main observations of Section \ref{six} are summarized in Table \ref{summary}.
\begin{table}
\renewcommand{\arraystretch}{0.9}
\caption{Summary of the observations on different CA schemes}
\label{summary} \centering
{\small
\begin{tabular}{ll|l|c}
\multicolumn{2}{c|}{\bf CA scheme}  &{\bf Performance ($R_{0.5}$, $R_{0.9}$, sum log-rate)}  &{\bf Complexity} \\
\hline
u-WBBA\  &SINR CA  &Lowest in all u-WBBA schemes in all aspects   &very low \\
         &CRE      &Slightly improved over SINR CA    &very low \\
         &CA-WBBA  &Significant improvement over SINR CA and CRE;     &medium \\
         &         &optimal with u-WBBA; in general poorer than p-WBBA     &       \\
\hline
p-WBBA\  &SINR CA       &No significant difference from others in $R_{0.5}$ and $R_{0.9}$;     &very low \\
         &              &low sum log-rate                            & \\
         &CRE           &Not guaranteed improvement over SINR CA     &very low \\
         &GAMS BONMIN   &Optimal in sum log-rate; good $R_{0.5}$ but poorest $R_{0.9}$     &highest \\
         &CA-WBBA       &Best $R_{0.9}$; good $R_{0.5}$ but no significant advantage       &medium high \\
         &              &over SINR CA; good sum log-rate, slightly lower than BONMIN       &  \\
         &Macro BS OFL  &Best $R_{0.5}$; good $R_{0.9}$ but no significant advantage       &low \\
         &              &over SINR CA; good sum log-rate, slightly lower than CA-WBBA          &  \\
\hline
\end{tabular}
}
%\vskip -6pt
\end{table}
In general, the schemes under p-WBBA outperform those under
u-WBBA, however with a higher system complexity.
The GAMS BONMIN solution, which gives the best objective function value, fails to achieve
good fairness among the MTs.
The proposed cell association schemes for the p-WBBA scenario, on the other hand, achieve more balanced QoS.
Specifically, the distributed joint CA-WBBA algorithm achieves the best fairness in terms of 90\% MT rate,
while the macro BS offloading algorithm gives the best median MT rate.

%%%%%%%%%%%%%%%%%%%%%%%%%%%%%%%%%%%%%%%%%%%%%%%%%%%%%%%%%%%%%%%%%%%%%%%%%%%%%%%%%%%%%%%%%%%%%%%%%%
\section{Conclusion}\label{seven}
We have studied joint downlink cell association and wireless backhaul bandwidth allocation
in a two-tier HetNet where the macro BS is equipped with large-scale antenna array,
and the small cells rely on in-band wireless links to the massive MIMO macro BS for backhauling.
A reverse-TDD and soft frequency reuse framework has been introduced for interference management
in the HetNet model.
With the wireless backhaul constraint,
the joint scheduling problem
for maximization of sum of log rates for MTs  has been shown to be a nonlinear mixed-integer programming problem.
Two wireless backhaul bandwidth allocation scenarios, which consider global (unified WBBA)
and local (per-small-cell WBBA) backhaul bandwidth allocation have been investigated.
By showing a separable concave objective function in the unified WBBA scenario,
a two-level hierarchical decomposition method for relaxed optimization has been considered,
and a distributed joint CA-WBBA algorithm has been proposed.
The algorithm has been extended to the per-small-cell WBBA scenario,
which achieves improved spectral efficiency and network utility.
Fast heuristic algorithms for cell association in the per-small-cell WBBA scenario has also been studied.
According to the numerical results, performances of the cell association algorithms are not sensitive
to the density of small cells in the two-tier HetNet.
An optimal load (i.e., number of associated MTs),
which is approximately equal to the number of antennas at the macro BS antenna array, has been observed.
It has also been shown that the cell range expansion strategy fails to provide good performance
when the wireless backhauling constraints for the small cells are considered for cell association.

\end{document}